\documentclass[12pt,showpacs,preprintnumbers,superscriptaddress,nofootinbib]{revtex4}
\usepackage{amssymb}
\usepackage{amsmath, graphicx}
\usepackage{dcolumn}
\usepackage{bm}
\usepackage{epstopdf}
\usepackage[utf8]{inputenc}

\input{tcilatex}
\begin{document}

\title{\boldmath The  $\Delta (27)$ flavor 3-3-1 model with neutral
leptons}
\author{V. V. Vien}
\email{wvienk16@gmail.com} \affiliation{Institute of Research and Development, Duy Tan University,\\ 182 Nguyen Van Linh, Da Nang City, Vietnam}
\affiliation{Department of Physics, Tay
Nguyen University, \\
567 Le Duan, Buon Ma Thuot, DakLak, Vietnam}
\author{A. E. C\'arcamo Hern\'andez}
\email{antonio.carcamo@usm.cl}
\affiliation{Universidad T\'{e}cnica Federico Santa Mar\'{\i}a\\
and Centro Cient\'{\i}fico-Tecnol\'{o}gico de Valpara\'{\i}so\\
Casilla 110-V, Valpara\'{\i}so, Chile,}
\author{H. N. Long}
\email{hnlong@iop.vast.ac.vn}
\affiliation{Institute of Physics, Vietnam Academy of Science and Technology, \\
10 Dao Tan, Ba Dinh, Hanoi, Vietnam}
\date{\today }

\begin{abstract}
We build the first 3-3-1 model based on the $\Delta (27)$ discrete group symmetry,
consistent with fermion masses and mixings. In the model under consideration, the  
neutrino masses are generated from a combination of type-I and type-II seesaw
mechanisms mediated by three heavy right-handed Majorana neutrinos and three
$SU(3)_{L}$ scalar antisextets, respectively. Furthermore, from the consistency of the leptonic mixing angles with their experimental values, we obtain a non-vanishing leptonic Dirac CP violating phase of $-\frac{\pi }{2}$. Our model features an effective Majorana neutrino mass parameter of neutrinoless double beta decay, with values $m_{\beta \beta }=$ 10 and 18 meV for the normal and the inverted neutrino mass hierarchies, respectively.

\end{abstract}

\maketitle

\newpage
\section{\label{intro} Introduction}

The discovery of the  $ 126$ GeV Higgs boson at the Large Hadron Collider (LHC) \cite{Aad:2012tfa,Chatrchyan:2012xdj}, has filled the vacancy of the Higgs boson
needed for the completion of the Standard Model (SM) at the Fermi scale and
has provided a confirmation for the mass generation mechanism of the weak
gauge bosons. Despite LHC experiments indicate that the decay modes of the
new scalar state are very close to the SM expectation, there is still room for new extra scalar
states. The search of these new scalar states will shed light on the
underlying theory behind Electroweak Symmetry Breaking (EWSB) and are the
priority of the LHC experiments. Furthermore, despite its great experimental
success, the SM has several unaddressed issues, such as, for example, the
observed charged fermion mass and quark mixing pattern, the tiny neutrino
masses and the sizeable leptonic mixing angles, which contrast with the
small quark mixing angles. The global fits of the available data
from the Daya Bay \cite{An:2012eh}, T2K \cite{Abe:2011sj}, MINOS
\cite{Adamson:2011qu}, Double CHOOZ \cite{Abe:2011fz} and RENO \cite{Ahn:2012nd}
neutrino oscillation experiments, provide constraints on the neutrino mass
squared splittings and mixing parameters \cite{Forero:2014bxa}. It is well
kwown that the charged fermion mass hierarchy spans over a range of five orders
of magnitude in the quark sector and a much wider range, which includes extra six orders of
magnitude, corresponding to the number of orders of
magnitude between the neutrino mass scale and the electron mass. The charged
fermion masses can be accommodated in the SM, at the price of having an
unnatural tuning among its different Yukawa couplings. Furthermore,
experiments with solar, atmospheric and reactor neutrinos
\cite{Agashe:2014kda,An:2012eh,Abe:2011sj,Adamson:2011qu,Abe:2011fz,Ahn:2012nd}
provide clear indications of neutrino oscillations, originated by
nonvanishing neutrino masses. All these unexplained issues suggest that new physics
have to be invoked to address the fermion puzzle of the SM.

\quad The unexplained flavour puzzle of the SM motivates to consider extensions of the SM that explain the fermion mass and mixing pattern. From the phenomenological point of view, one can assume Yukawa textures \cite{Fritzsch:1977za,Fukuyama:1997ky,Du:1992iy,Barbieri:1994kw,Peccei:1995fg,Fritzsch:1999ee,Roberts:2001zy,Nishiura:2002ei,deMedeirosVarzielas:2005ax,Carcamo:2006dp,Kajiyama:2007gx,CarcamoHernandez:2010im,Branco:2010tx,Leser:2011fz,Gupta:2012dma,Hernandez:2013mcf,Pas:2014bra,Hernandez:2014hka,Hernandez:2014zsa,Nishiura:2014psa,Frank:2014aca,Sinha:2015ooa,Nishiura:2015qia,Gautam:2015kya,Pas:2015hca} to explain some features of the fermion mass hierarchy. Discrete flavor groups provide a very promising approach to address the
flavour puzzle, and been extensively used in several models to explain the
prevailing pattern of fermion masses and mixings (see Refs. \cite{Ishimori:2010au,Altarelli:2010gt,King:2013eh, King:2014nza} for recent
reviews on flavor symmetries). Models with spontaneously broken flavor
symmetries may also produce hierarchical mass structures. Recently, discrete
groups such as $A_{4}$\cite{Ma:2001dn,He:2006dk,Chen:2009um,Dong:2010gk,Ahn:2012tv,Memenga:2013vc,Felipe:2013vwa,Varzielas:2012ai,Ishimori:2012fg,King:2013hj,Hernandez:2013dta,Babu:2002dz,Altarelli:2005yx,Morisi:2013eca,Altarelli:2005yp,Kadosh:2010rm,Kadosh:2013nra,delAguila:2010vg,Campos:2014lla,Vien:2014pta,Hernandez:2015tna,Nishi:2016jqg}, $S_{3}$ \cite{Chen:2004rr,Dong:2011vb,Bhattacharyya:2010hp,Dias:2012bh,Meloni:2012ci,Canales:2013cga,Ma:2013zca,Kajiyama:2013sza,Hernandez:2013hea,Ma:2014qra,Hernandez:2014vta,Hernandez:2014lpa,Hernandez:2015dga,Hernandez:2015zeh,Hernandez:2016rbi,Hernandez:2015hrt}, $S_{4}$ \cite{Mohapatra:2012tb,Varzielas:2012pa,Ding:2013hpa,Ishimori:2010fs,Ding:2013eca,Hagedorn:2011un,Campos:2014zaa,Dong:2010zu,VanVien:2015xha}, $D_4$ \cite{Frampton:1994rk,Grimus:2003kq,Grimus:2004rj,Frigerio:2004jg,Babu:2004tn,Adulpravitchai:2008yp,Ishimori:2008gp,Hagedorn:2010mq,Meloni:2011cc,Vien:2013zra}, $T_7$ \cite{Luhn:2007sy,Hagedorn:2008bc,Cao:2010mp,Luhn:2012bc,Kajiyama:2013lja,Bonilla:2014xla,Vien:2014gza,Vien:2015koa,Hernandez:2015cra,Arbelaez:2015toa}, $T_{13}$ \cite{Ding:2011qt,Hartmann:2011dn,Hartmann:2011pq,Kajiyama:2010sb}, $T^{\prime }$ \cite{Aranda:2000tm,Aranda:2007dp,Chen:2007afa,Frampton:2008bz,Eby:2011ph,Frampton:2013lva} and $\Delta(27)$ \cite{Varzielas:2012nn,Bhattacharyya:2012pi,Ma:2013xqa,Nishi:2013jqa,Varzielas:2013sla,Aranda:2013gga,Varzielas:2015aua,Chen:2015jta} have been implemented in extensions of the SM to explain the prevailing fermion mass and mixing pattern.

\quad Besides that, another unaswered issue in particle physics is the
existence of three families of fermions at low energies. The origin of
the family structure of the fermions can be addressed in family dependent
models where a symmetry distinguish fermions of different families. This issue can be explained by the models based on the $SU(3)_{C}\otimes SU(3)_{L}\otimes U(1)_{X}$ gauge symmetry, also called 3-3-1 models, which include a family non-universal $U(1)_{X}$ symmetry
\cite{Georgi:1978bv,Valle:1983dk,Pisano:1991ee,Foot:1992rh,Frampton:1992wt,Ng:1992st,Duong:1993zn,Hoang:1996gi,Hoang:1995vq,Foot:1994ym,Diaz:2003dk,Diaz:2004fs,Dias:2004dc,Ochoa:2005ch,CarcamoHernandez:2005ka,Alvarado:2012xi,Catano:2012kw,Hernandez:2013mcf,Hernandez:2014lpa,Vien:2014pta,Hernandez:2014vta,Boucenna:2014ela,Boucenna:2014dia,Vien:2014gza,Phong:2014ofa,Boucenna:2015zwa,Hernandez:2015cra,DeConto:2015eia,Correia:2015tra,Hernandez:2015tna,Okada:2015bxa,Long:2015gca,Long:2015qza,Binh:2015cba,Hue:2015fbb,Boucenna:2015pav,Hernandez:2015ywg,Dong:2015dxw,Pal:1994ba,Mizukoshi:2010ky}. These models have several phenomenological advantages. Firstly, the
three family structure in the fermion sector can be explained in the 3-3-1
models from the chiral anomaly cancellation and asymptotic freedom in
QCD \cite{331Pisano:1992, 331Frampton:1992, 331Foot:1993}. Secondly, the fact that the third
family is treated under a different
representation, can explain the large mass difference between the heaviest quark family and t
he two lighter ones. Finally, these models contain a
natural Peccei-Quinn symmetry, necessary to solve the strong-CP problem \cite{Pal:1994ba}.
Furthermore, the 331 models with sterile neutrinos have weakly
interacting massive fermionic dark matter candidates~\cite{Mizukoshi:2010ky}.

\quad In the 3-3-1 models, the $SU(3)_{L}\otimes U(1)_{X}$ symmetry is
broken down to the SM electroweak group $SU(2)_{L}\otimes U(1)_{Y}$ by one
heavy $SU(3)_L$ triplet field that gets a Vacuum Expectation Value (VEV) at
high energy scale $v _{\chi }$, thus giving masses to non
SM fermions and gauge bosons, while the Electroweak Symmetry Breaking is
triggered by the remaining lighter triplets as well as by $SU(3)_L$
antisextets in some version of the model, with VEVs at the electroweak scale
$\upsilon _{\rho }$ and $\upsilon _{\eta }$, thus providing masses for SM
fermions and gauge bosons \cite{Hernandez:2013mcf}.

In this paper we propose a 3-3-1 model based on the $\mathrm{SU}(3)_{C}\otimes
\mathrm{SU}(3)_{L}\otimes \mathrm{U}(1)_{X}\otimes \mathrm{U}(1)_{\mathcal{L}%
}\otimes \Delta (27)$ symmetry consistent with fermion masses
and mixings. Our model is the first 331 model based on the $\Delta (27)$
family symmetry, proposed in the literature\footnote{In this scenario, only one flavor
 symmetry $\Delta (27)$ is added.}. Our model also includes a new
 $\mathrm{U}(1)_{\mathcal{L}}$ that allows us to treat the quark, charged
lepton and neutrino sector independently. The light active
neutrino masses arise from a combination of type I and type II seesaw
mechanisms mediated by three heavy right handed Majorana neutrinos and three
$SU(3)_{L}$ scalar antisextets, respectively. The content of this paper goes
as follows. In Sec. \ref{model} we explain some theoretical aspects of our
331 model. The charged fermion sector is discussed in Sec \ref{clep}. In Sec. %
\ref{neutrino} we focus on the discussion of the neutrino sector as well
as in lepton masses and mixing and give our corresponding results. In Sec. %
\ref{quark}, we discuss the implications of our model in the quark sector.
Conclusions are given in Sec. \ref{conclusion}. In the appendices
we present several technical details: Appendices \ref{Delta27g} and %
\ref{Delta27repre3} give a detailed description of the $\Delta (27)$ group
and the matrices of the 3 representation of $\Delta (27)$, respectively.
The Appendix \ref{D27breaking3} provides the breaking
patterns of $\Delta (27)$ by triplets. We prefer to use the notation $3^*$ for a $SU(3)$ anti-triplet
and $\bar{\underline{3}}$ for a $\Delta (27)$ anti-triplet, i. e., all $\Delta (27)$ representations appear with a bar underneath, and the anti-triplets appear also with a bar on top.

\section{The model \label{model}}

The symmetry group of the model under consideration is
\begin{equation*}
G=\mathrm{SU}(3)_{C}\otimes \mathrm{SU}(3)_{L}\otimes \mathrm{U}%
(1)_{X}\otimes \mathrm{U}(1)_{\mathcal{L}}\otimes \Delta (27),
\end{equation*}%
where the electroweak factor $\mathrm{SU}(3)_{L}\otimes \mathrm{U}(1)_{X}$
is extended from those of the SM, and the strong interaction sector
is retained. Lets us note that the gauge symmetry of the $331$ model is
supplemented by the $\mathrm{U}(1)_{\mathcal{L}}$ global and $\Delta (27)$
symmetries. Each lepton family includes a new neutral fermion $(N_{R})$ with
vanishing lepton number $L(N_{R})=0$ arranged under the $\mathrm{SU}(3)_{L}$
symmetry as a triplet $(\nu _{L},l_{L},N_{R}^{c})$ and a singlet $l_{R}$.
The residual electric charge operator $Q$ is therefore related to the
generators of the gauge symmetry by \cite{Dong:2010zu}
\begin{equation*}
Q=T_{3}-\frac{1}{\sqrt{3}}T_{8}+X,
\end{equation*}%
where $T_{a}$ $(a=1,2,...,8)$ are $\mathrm{SU}(3)_{L}$ charges with $\mathrm{%
Tr}T_{a}T_{b}=\frac{1}{2}\delta _{ab}$ and $X$ is the $\mathrm{U}(1)_{X}$
charge. This means that the model under consideration does not contain
exotic electric charges in the fundamental fermion, scalar and adjoint gauge
boson representations. Since particles with different lepton number are put
in $\mathrm{SU}(3)_{L}$ triplets, it is better to work with a new conserved
charge $\mathcal{L}$ commuting with the gauge symmetry and related to the
ordinary lepton number by diagonal matrices \cite{Dong:2010zu,Chang:2006aa}
\begin{equation*}
L=\frac{2}{\sqrt{3}}T_{8}+\mathcal{L}.
\end{equation*}

The lepton charge arranged in this way, i.e. $L(N_R)=0$, is in order to
prevent unwanted interactions due to $\mathrm{U}(1)_\mathcal{L}$ symmetry
and breaking due to the lepton parity to obtain the consistent lepton and
quark spectra. By this embedding, exotic quarks $U, D$ as well as new
non-Hermitian gauge bosons $X^0$, $Y^\pm$ possess lepton charges as of the
ordinary leptons: $L(D)=-L(U)=L(X^0)=L(Y^{-})=1$.

The fermion content and the scalar fields of the model are summarized in Tab. \ref{Fercon}.
\begin{table}[ht]
\begin{center}
\caption{\label{Fercon} The fermion content of the model.}
\begin{tabular}{|c||c|c|c|c|c|c|c|c||c|c|c|c|c|c|c|c|}
\hline
        Fields & $\psi_{1,2,3L}$ &$l_{1,2, 3R}$&\,\,$Q_{1,2L}$&\,\,$Q_{3L}$\,\,&\,\,$u_{R}$\,\,&\,\,$d_{R}$\,\,&\,\,$U_R$\,\,&\,\,$D_{1,2R}$&$\phi$&$\sigma$&$\rho$&$\eta$&$\chi$\\ \hline\hline
$\mathrm{SU}(3)_L$  & $3$ &$1$& $3^*$&$3$& $1$&$1$&$1$&$1$&$3$&$6^*$&$3$&$3$&$3$  \\ \hline
$\mathrm{U}(1)_X$ & $-\frac{1}{3}$  &$-1$& $0$&$\frac{1}{3}$& $\frac{2}{3}$& $-\frac{1}{3}$&$\frac{2}{3}$&$-\frac{1}{3}$&$\frac{2}{3}$&$\frac{2}{3}$&$\frac{2}{3}$&$-\frac{1}{3}$&$-\frac{1}{3}$  \\\hline
$\mathrm{U}(1)_{\mathcal{L}}$ & $\frac{2}{3}$  &$1$&  $\frac{1}{3}$&$-\frac{1}{3}$& $0$& $0$&$-1$& $1$&$-\frac{1}{3}$&$-\frac{4}{3}$&$-\frac{4}{3}$&$-\frac{1}{3}$&$\frac{2}{3}$ \\\hline
$\underline{\Delta}(27)$&  $\underline{3}$  &$\underline{1}_1, \underline{1}_2, \underline{1}_3$& $\underline{1}_{1,2}$& $\underline{1}_{3}$&$\underline{3}$&$\underline{\bar{3}}$&$\underline{1}_{2}$&$\underline{1}_{1,3}$&$\underline{3}$&$\underline{3}$&$\underline{3}$&$\underline{\bar{3}}$&$\underline{1}_1$   \\ \hline
\end{tabular}
\end{center}
\end{table}

As we will see in the next sections, the $\mathrm{U}(1)_{X}$ and $\mathrm{U}%
(1)_{\mathcal{L}}$ charge assignments for the fermion sector, enforce to
have different scalar fields in the quark, charged leptons and neutrino
Yukawa interactions. Consequently the $\mathrm{U}(1)_{X}$ and $\mathrm{U}%
(1)_{\mathcal{L}}$ symmetries help to treat the charged lepton, neutrino and
quark sectors independently.

\subsection{Charged -lepton sector\label{clep}}

Since left handed $SU\left( 3\right) _{L}$ lepton triplets are unified in a $%
\Delta \left( 27\right) $ triplet, to generate charged lepton masses, we
need three $SU(3)_{L}$ Higgs triplets grouped in a $\underline{3}$ under $%
\Delta (27)$ given in Tab. \ref{Fercon} . The $G$ assignments of the scalar fields participating in
charged lepton Yukawa interactions are:
\begin{equation}
\phi =\left( \phi _{1},\phi _{2},\phi _{3}\right), %\sim \lbrack 3,2/3,-1/3,%\underline{3}]
\hspace*{0.5cm}\phi _{i}=\left( \phi _{i1}^{+}\,\,,
\phi _{i2}^{0} \,\,,
\phi _{i3}^{+}\right)^T ,\hspace*{0.5cm} i=1,2,3.
\end{equation}%
The Yukawa interactions for charged leptons are
\begin{eqnarray}
-\mathcal{L}_{l} &=&h_{1}(\bar{\psi}_{L}\phi )_{\underline{1}%
_{1}}l_{1R}+h_{2}(\bar{\psi}_{L}\phi )_{\underline{1}_{3}}l_{2R}+h_{3}(\bar{%
\psi}_{L}\phi )_{\underline{1}_{2}}l_{3R}+H.c  \notag \\
&=&h_{1}(\bar{\psi}_{1L}\phi _{1}+\bar{\psi}_{2L}\phi _{2}+\bar{\psi}%
_{3L}\phi _{3})_{\underline{1}_{1}}l_{1R}  \notag \\
&+&h_{2}(\bar{\psi}_{1L}\phi _{1}+\omega ^{2}\bar{\psi}_{2L}\phi _{2}+\omega
\bar{\psi}_{3L}\phi _{3})_{\underline{1}_{1}}l_{2R}  \notag \\
&+&h_{3}(\bar{\psi}_{1L}\phi _{1}+\omega \bar{\psi}_{2L}\phi _{2}+\omega ^{2}%
\bar{\psi}_{3L}\phi _{3})_{\underline{1}_{1}}l_{3R}+H.c.  \label{Lclep0}
\end{eqnarray}%
To obtain a realistic lepton spectrum, we suppose that in charged lepton
sector $\Delta (27)$ is broken down to $\{\mathrm{Identity}\}$,
i.e, it is completely broken. This can be achieved with the VEV alignment $%
\langle \phi \rangle =(\langle \phi _{1}\rangle ,\langle \phi _{2}\rangle
,\langle \phi _{3}\rangle )$ under $\Delta (27)$, where $\langle \phi
_{1}\rangle \neq \langle \phi _{2}\rangle \neq \langle \phi _{3}\rangle $,
and
\begin{equation}
\langle \phi _{i}\rangle =\left( 0\,\,\,\,\,v_{i}\,\,\,\,\,0\right)
^{T},\,\,\,(i=1,2,3).
\end{equation}%
Under this alignment, the mass Lagrangian for the charged leptons reads
\begin{equation}
\mathcal{L}_{l}^{\mathrm{mass}}=-(\bar{l}_{1L},\bar{l}_{2L},\bar{l}%
_{3L})M_{l}(l_{1R},l_{2R},l_{3R})^{T}+H.c,
\end{equation}%
where
\begin{equation}
M_{l}=\left(
\begin{array}{ccc}
h_{1}v_{1} & h_{2}v_{1} & h_{3}v_{2} \\
h_{1}v_{2} & \hspace*{0.5cm}\omega ^{2}h_{2}v_{2} & \hspace*{0.5cm}\,\omega
h_{3}v_{2} \\
h_{1}v_{3} & \hspace*{0.5cm}\,\,\omega h_{2}v_{3} & \,\,\,\,\,\omega
^{2}h_{3}v_{3}%
\end{array}%
\right) .  \label{Mlep}
\end{equation}%
As will shown in section \ref{neutrino}, in the case $v_{1}=v_{2}=v_{3}=v$,
i.e, $\Delta (27)$ is broken into $Z_{3}$ group which consisting of the
elements \{$1,b,b^{2}$\}, the charged lepton matrix $M_{l}$ in Eq. (\ref%
{Mlep}) is diagonalized by the matrix
\begin{equation}
U_{0L}=\frac{1}{\sqrt{3}}\left(
\begin{array}{ccc}
1 & 1 & 1 \\
1 & \omega ^{2} & \omega  \\
1 & \omega  & \omega ^{2}%
\end{array}%
\right) ,  \label{U0clep}
\end{equation}%
and the exact tri-bimaximal mixing form will obtained. For a detailed study
of this problem, the reader can see Ref. \cite{Vien:2014gza}.

As we know, the realistic lepton mixing form is a small deviation from
tri-bimaximal form \cite{Agashe:2014kda} . This can be achieved with a small
difference between $v_{2},v_{3}$ and $v_{1}$. Therefore we can separate $v_{2},v_{3}$ into two parts, the first is equal to $v_{1}\equiv v$, the
second is responsible for the deviation,
\begin{equation}
v_{1}=v,\,\,v_{2}=v(1+\varepsilon _{2}),\,\,v_{3}=v(1+\varepsilon
_{3}),\,\,\varepsilon _{2,3}\ll 1,
\end{equation}%
and the matrix $M_{l}$ in (\ref{Mlep}) becomes
\begin{eqnarray}
M_{l} &=&\left(
\begin{array}{ccc}
h_{1}v & h_{2}v & h_{3}v \\
h_{1}v(1+\varepsilon _{2}) & \omega ^{2}h_{2}v(1+\varepsilon _{2}) &
\,\,\omega h_{3}v(1+\varepsilon _{2}) \\
h_{1}v(1+\varepsilon _{3}) & \,\,\,\omega h_{2}v(1+\varepsilon _{3}) &
\,\,\omega ^{2}h_{3}v(1+\varepsilon _{3})%
\end{array}%
\right)   \notag \\
&\equiv &v\left(
\begin{array}{ccc}
1 & 0 & 0 \\
0 & \,\,\,1+\varepsilon _{2} & 0 \\
0 & 0 & 1+\varepsilon _{3}%
\end{array}%
\right) \left(
\begin{array}{ccc}
1 & 1 & 1 \\
1 & \omega ^{2} & \omega  \\
1 & \omega  & \omega ^{2}%
\end{array}%
\right) \left(
\begin{array}{ccc}
h_{1} & 0 & 0 \\
0 & h_{2} & 0 \\
0 & 0 & h_{3}%
\end{array}%
\right) .  \label{Mlep1}
\end{eqnarray}%
The matrix $M_{l}$ in Eq. (\ref{Mlep1}) can be diagonalized by two steps as
follows.\newline
Firstly, we denote

\begin{equation}
M_{l}^{\prime }=U_{0L}^{+}M_{l}=\frac{v}{\sqrt{3}}\left(
\begin{array}{ccc}
(3+\varepsilon _{2}+\varepsilon _{3})h_{1} & (\omega ^{2}\varepsilon
_{2}+\omega \varepsilon _{3})h_{2} & (\omega \varepsilon _{2}+\omega
^{2}\varepsilon _{3})h_{3}\notag \\
(\omega \varepsilon _{2}+\omega ^{2}\varepsilon _{3})h_{1} & (3+\varepsilon
_{2}+\varepsilon _{3})h_{2} & (\omega ^{2}\varepsilon _{2}+\omega
\varepsilon _{3})h_{3}\notag \\
\omega ^{2}\varepsilon _{2}+\omega \varepsilon _{3})h_{1} & (\omega
\varepsilon _{2}+\omega ^{2}\varepsilon _{3})h_{2} & (3+\varepsilon
_{2}+\varepsilon _{3})h_{3}%
\end{array}%
\right),
\label{Mlepp}
\end{equation}%
Secondly, the matrix $M_{l}^{\prime }$ in Eq. (\ref{Mlepp}) is diagonalized
by
\begin{equation}
U_{L}^{+}M_{l}^{\prime }\equiv U_{L}^{+}U_{0L}^{+}M_{l}=\mathrm{diag}%
(m_{e},m_{\mu },m_{\tau }),  \label{Mleppv}
\end{equation}%
where
\begin{equation}
m_{e}=Y_{l}h_{1}v,\,\,m_{\mu }=Y_{l}h_{2}v,\,\,m_{\tau }=Y_{l}h_{3}v,
\label{memutau}
\end{equation}%
with
\begin{equation*}
Y_{l}=\frac{3\sqrt{3}(1+\varepsilon _{3})[\varepsilon _{3}(\varepsilon
_{3}+\varepsilon -4)-4]}{(2+\varepsilon _{3})[\varepsilon _{3}(\varepsilon
_{3}+\varepsilon -6)-6]},
\end{equation*}%
and
\begin{equation*}
\varepsilon =\sqrt{\varepsilon _{3}^{2}-12(\varepsilon _{3}+1)}.
\end{equation*}%
The matrix that diagonalize $M_{l}^{\prime }$ in (\ref{Mlepp}) takes the
form:
\begin{equation}
U_{L}=\left(
\begin{array}{ccc}
1 & U_{12}^{l} & U_{13}^{l} \\
U_{13}^{l} & 1 & U_{12}^{l} \\
U_{12}^{l} & U_{13}^{l} & 1%
\end{array}%
\right) ,\hspace*{0.5cm}U_{R}=1  \label{Uclep}
\end{equation}%
where
\begin{eqnarray}
U_{12}^{l} &=&\frac{\varepsilon _{3}\left\{ 6-2i\sqrt{3}-(1+i\sqrt{3}%
)\varepsilon +\varepsilon _{3}[7-i\sqrt{3}+(1-i\sqrt{3})(\varepsilon
-\varepsilon _{3})]\right\} }{2(2+\varepsilon _{3})[-6+\varepsilon
_{3}^{2}-\varepsilon _{3}(6+\varepsilon )]},  \notag \\
U_{13}^{l} &=&\frac{\varepsilon _{3}\left\{ 6+2i\sqrt{3}-(1-i\sqrt{3}%
)\varepsilon +\varepsilon _{3}[7+i\sqrt{3}+(1+i\sqrt{3})(\varepsilon
-\varepsilon _{3})]\right\} }{2(2+\varepsilon _{3})[-6+\varepsilon
_{3}^{2}-\varepsilon _{3}(6+\varepsilon )]},\hspace*{0.5cm}  \label{u12u13l}
\end{eqnarray}%
To get the results in Eq.(\ref{u12u13l}) we have used the following
relations
\begin{equation*}
\varepsilon _{2}=\frac{\varepsilon _{3}(2-\varepsilon _{3}-\varepsilon )}{%
2(2+\varepsilon _{3})},\hspace*{0.5cm}\varepsilon _{2}^{\ast }=\frac{%
\varepsilon _{3}\left( -2-3\varepsilon _{3}+\varepsilon \right) }{%
2(1+\varepsilon _{3})(2+\varepsilon _{3})},\hspace*{0.5cm}\varepsilon
_{3}^{\ast }=\frac{1}{1+\varepsilon _{3}}-1,
\end{equation*}%
which are obtained from the unitary condition of $U_{L}$.\newline
The left- and right- handed mixing matrices in charged lepton sector are
given by:
\begin{equation}
U_{L}^{\prime }=U_{0L}.U_{L}=\left(
\begin{array}{ccc}
\alpha _{1} & \alpha _{1} & \alpha _{1} \\
\alpha _{2} & \omega ^{2}\alpha _{2} & \omega \alpha _{2} \\
\alpha _{3} & \omega \alpha _{3} & \omega ^{2}\alpha _{3}%
\end{array}%
\right) ,\,\,U_{R}^{\prime }=1,  \label{Uclepp}
\end{equation}%
where\footnote{With the value of $\epsilon$ obtainted in Eq.(\ref{ep31}), $\left|\alpha _{1}\right|\simeq\left|\alpha _{2}\right|\simeq\left|\alpha _{3}\right|=0.577\simeq 1/\sqrt{3}$.}
\begin{equation}
\alpha _{1}=\frac{\sqrt{3}\left[ \varepsilon _{3}^{2}-\varepsilon
_{3}(\varepsilon +4)-4\right] }{(2+\varepsilon _{3})[\varepsilon
_{3}^{2}-\varepsilon _{3}(\varepsilon +6)-6]},\hspace*{0.5cm}\alpha _{2}=%
\frac{2\sqrt{3}(1+\varepsilon _{3})}{6-\varepsilon _{3}^{2}+\varepsilon
_{3}(6+\varepsilon )},\hspace*{0.5cm}\alpha _{3}=(1+\varepsilon _{3})\alpha
_{1}.
\label{al123}
\end{equation}%
In the case $\varepsilon _{3}=0$ it folows that $\varepsilon _{2}^{\ast
}=\varepsilon _{2}=\varepsilon _{3}^{\ast }=0$, $U_{L}=1$ and the lepton
mixing $U_{L}^{\prime }$ in Eq. (\ref{Uclepp}) reduces to Tri-bimaximal form ($U_{HPS}$) \cite{HarrisonPS:2002} which is
ruled out by the recent data \cite{Agashe:2014kda}. In general $\varepsilon
_{2,3}\neq 0$ (but small) so $\alpha _{i}\,\,(i=1,2,3)$ in Eq. (\ref{al123})
are a little different to each other and different from $\frac{1}{\sqrt{3}}$%
. Consequently, the lepton mixing $U_{L}^{\prime }$ in Eq. (\ref{Uclepp})
differs to $U_{HPS}$ and can lead to the realistic lepton mixing with
non-zero $\theta _{13}$ as represented in Sec.\ref{neutrino}. This is one of
the striking results of the model under consideration.

Taking into account of the discovery of the long-awaited Higgs boson at
around 125 GeV by ATLAS\cite{Aad:2012tfa} and CMS \cite{Chatrchyan:2012xdj},
we can choose\footnote{%
In the SM, the Higgs VEV is equal to 246 $\mathrm{GeV}$, fixed by the $W$
boson mass $m_{W}^{2}=\frac{g^{2}}{4}v_{weak}^{2}$, and in the model under
consideration, $M_{W}^{2}\simeq \frac{g^{2}}{2}\left(3u^{2}+3v^{2}\right) $. Therefore, we can identify $v_{weak}^{2}=6(u^{2}+v^{2})=(246\,\mathrm{GeV})^{2}$ and then obtain $u\sim v\simeq 71\,\mathrm{GeV}$.}  $v=100\,\mathrm{GeV}$ for
its scale.
From (\ref{memutau}), the charged
lepton Yukawa couplings $h_{1,2,3}$ relate to their masses as follows:
\begin{equation}
h_{1}=m_{e}/Y_{l}v,\,\,h_{2}=m_{\mu }/Y_{l}v,\,\,h_{3}=m_{\tau }/Y_{l}v.  \label{h123v}
\end{equation}%
The best fit values for the charged lepton masses are given in Ref.
\cite{Agashe:2014kda}:
\begin{equation}
m_{e}\simeq 0.511\,\text{MeV},\,\,m_{\mu }\simeq 105.66\,\text{MeV}%
,\,\,m_{\tau }\simeq 1776.82\,\text{MeV}\mathbf{.}  \label{memutaexpv}
\end{equation}%
With the help of Eqs. (\ref{memutaexpv}) and (\ref{h123v}) we get $%
\frac{h_{1}}{h_{2}}\simeq 0.0048,\,\,\frac{h_{1}}{h_{3}}\simeq 0.00029$ and $%
\frac{h_{2}}{h_{3}}=0.0595$, i.e, $h_{1}\ll h_{2}\ll h_{3}$ for $\varepsilon
_{3}$ is arbitrary. As will be shown in Sec.\ref{neutrino}, from the experimental constrains on lepton mixing \cite{GonzalezGarcia:2014sz}, we obtain a solution in Eq. (\ref{ep31}). With this solution,
we get
\begin{equation*}
h_{1}=2.96671\times 10^{-6},\,\,h_{2}=6.13429\times
10^{-4},\,\,h_{3}=1.03157\times 10^{-2}.
\end{equation*}%
We note that the mass hierarchy of the charged leptons are well separated by
only one Higgs triplet $\phi $ of $\Delta (27)$, and this is one of the good features of the $\Delta (27)$ group.

\subsection{Neutrino masses and mixings\label{neutrino}}
The neutrino masses arise from the coupling of $\bar{\psi}_{L}^{c}\psi _{L}$
to scalars, where $\bar{\psi}_{L}^{c}\psi _{L}$ transforms as $3^{\ast
}\oplus 6$ under $\mathrm{SU}(3)_{L}$ and $\underline{\bar{3}}\oplus \underline{\bar{3}}\oplus \underline{\bar{3}}$ under $\Delta (27)$. It is worth noting that under the $\Delta (27)$
group, $\underline{3}\otimes \underline{3}\otimes \underline{3}$ has three
invariants. Consequently, to build neutrino Yukawa terms invariant under the
symmetries of the model, that give rise to light active neutrino masses via
type I and type II seesaw mechanisms, we enlarge the scalar sector of the $%
331$ model by introducing three $SU(3)_{L}$ scalar antisextets, namely $%
\sigma _{i}$ ($i=1,2,3$) as well as extra three $SU(3)_{L}$ scalar triplets,
denoted as $\rho _{i}$\ ($i=1,2,3$)\ grouped in $\Delta \left( 27\right) $
triplets as given in Tab.\ref{Fercon}. The scalar fields participating in the neutrino Yukawa
interactions have the following assignments under the $\mathrm{SU}(3)_{L}\otimes \mathrm{U}(1)_{X}\otimes \mathrm{U}%
(1)_{\mathcal{L}}\otimes \Delta (27)$ group:
\begin{eqnarray}
\sigma  &=&\left( \sigma _{1},\sigma _{2},\sigma _{3}\right),% \sim \lbrack 6^{\ast },2/3,-4/3,\underline{3}]
\hspace*{0.5cm}\sigma _{i}=\left(
\begin{array}{ccc}
\sigma _{11}^{0} & \sigma _{12}^{+} & \sigma _{13}^{0} \\
\sigma _{12}^{+} & \sigma _{22}^{++} & \sigma _{23}^{+} \\
\sigma _{13}^{0} & \sigma _{23}^{+} & \sigma _{33}^{0}%
\end{array}%
\right) _{i},\hspace*{0.5cm}i=1,2,3,  \label{sigma} \\
\rho  &=&\left( \rho _{1},\rho _{2},\rho _{3}\right),% \sim \lbrack 3,2/3,-4/3,%\underline{3}]
\hspace*{0.5cm}\rho _{i}=\left( \rho _{i1}^{+}\hspace*{0.5cm}%
\rho _{i2}^{0}\hspace*{0.5cm}\rho _{i3}^{+}\right) ^{T}.  \notag
\end{eqnarray}
Furthermore, we assume the following VEV patterns for the $\Delta
\left( 27\right) $ scalar triplets $\sigma $ and $\rho $:

\begin{equation*}
\langle \sigma \rangle =(\langle \sigma _{1}\rangle ,0,0),\hspace*{0.5cm}%
\hspace*{0.5cm}\langle \rho \rangle =(0,0,\langle \rho _{3}\rangle )
\end{equation*}%
where
\begin{equation*}
\langle \sigma _{1}\rangle =\left(
\begin{array}{ccc}
\lambda _{\sigma } & 0 & v_{\sigma } \\
0 & 0 & 0 \\
v_{\sigma } & 0 & \Lambda _{\sigma }%
\end{array}%
\right) ,\hspace*{0.5cm}\hspace*{0.5cm}\langle \rho _{3}\rangle =(0,v_{\rho
},0)^{T},
\end{equation*}%
i.e, $\Delta \left( 27\right) $ is broken into $Z_{3}$ groups which
consisting of the elements $\{e,aa^{\prime },(aa^{\prime})^2\}$ and \{$e,a^{\prime }, a^{\prime 2}$\} by $\sigma $ and $\rho $, respectively.

The neutrino Yukawa interactions invariant under the symmetries of the model
are given by \footnote{%
The following terms are invariant under the symmetries of the model: $(\bar{\psi}_{L}^{c}\sigma )_{\bar{3}}\psi
_{L}=\bar{\psi}_{2L}^{c}\sigma _{3}\psi _{1L}-\bar{\psi}_{3L}^{c}\sigma
_{2}\psi _{1L}+\bar{\psi}_{3L}^{c}\sigma _{1}\psi _{2L}-\bar{\psi}%
_{1L}^{c}\sigma _{3}\psi _{2L}+\bar{\psi}_{1L}^{c}\sigma _{2}\psi _{3L}-\bar{%
\psi}_{2L}^{c}\sigma _{1}\psi _{3L},(\bar{\psi}_{L}^{c}\rho )_{\bar{3}}\psi
_{L}=\bar{\psi}_{1L}^{c}\rho _{1}\psi _{1L}+\bar{\psi}_{2L}^{c}\rho _{2}\psi
_{2L}+\bar{\psi}_{3L}^{c}\rho _{3}\psi _{3L}$, and $(\bar{\psi}_{L}^{c}\rho
)_{\bar{3}}\psi _{L}=\bar{\psi}_{2L}^{c}\rho _{3}\psi _{1L}+\bar{\psi}%
_{3L}^{c}\rho _{2}\psi _{1L}+\bar{\psi}_{3L}^{c}\rho _{1}\psi _{2L}+\bar{\psi%
}_{1L}^{c}\rho _{3}\psi _{2L}+\bar{\psi}_{1L}^{c}\rho _{2}\psi _{3L}+\bar{%
\psi}_{2L}^{c}\rho _{1}\psi _{3L}$ but they are all vanish , i.e., they have no contribution to the neutrino mass matrices $M_{L,D,R}$.}:
\begin{eqnarray}
-\mathcal{L}_{\nu } &=&\frac{x}{2}(\bar{\psi}_{L}^{c}\sigma )_{\bar{3}}\psi
_{L}+\frac{y}{2}(\bar{\psi}_{L}^{c}\sigma )_{\bar{3}}\psi _{L}+\frac{z}{2}(%
\bar{\psi}_{L}^{c}\rho )_{\bar{3}}\psi _{L}+H.c  \notag \\
&=&\frac{x}{2}(\bar{\psi}_{1L}^{c}\sigma _{1}\psi _{1L}+\bar{\psi}%
_{2L}^{c}\sigma _{2}\psi _{2L}+\bar{\psi}_{3L}^{c}\sigma _{3}\psi _{3L}) +\frac{y}{2}\left( \bar{\psi}_{2L}^{c}\sigma _{3}\psi _{1L}+\bar{\psi}%
_{3L}^{c}\sigma _{2}\psi _{1L}+\bar{\psi}_{3L}^{c}\sigma _{1}\psi _{2L}\right.  \notag
\\
&&\left. \,\,\,\,
+\bar{%
\psi}_{1L}^{c}\sigma _{3}\psi _{2L}+\bar{\psi}_{1L}^{c}\sigma _{2}\psi _{3L}+%
\bar{\psi}_{2L}^{c}\sigma _{1}\psi _{3L}\right)
+\frac{z}{2}\left( \bar{\psi}_{2L}^{c}\rho _{3}\psi _{1L}-\bar{\psi}%
_{3L}^{c}\rho _{2}\psi _{1L}+\bar{\psi}_{3L}^{c}\rho _{1}\psi _{2L}\right.  \notag
\\
&&\left. \,\,\,\,-\bar{\psi%
}_{1L}^{c}\rho _{3}\psi _{2L}+\bar{\psi}_{1L}^{c}\rho _{2}\psi _{3L}-\bar{%
\psi}_{2L}^{c}\rho _{1}\psi _{3L}\right) +H.c,  \label{Lny}
\end{eqnarray}
Then, it follows that the neutrino mass terms are
\begin{eqnarray}
-\mathcal{L}_{\nu }^{mass} &=&\frac{1}{2}x[\lambda _{\sigma }\bar{\nu}%
_{1L}^{c}\nu _{1L}+v_{\sigma }\bar{N}_{1R}\nu _{1L}+v_{\sigma }\bar{\nu}%
_{1L}^{c}N_{1R}^{c}+\Lambda _{\sigma }\bar{N}_{1R}N_{1R}^{c}]  \notag \\
&+&\frac{1}{2}y\left[ \lambda _{\sigma }\bar{\nu}_{2L}^{c}\nu
_{3L}+v_{\sigma }\bar{N}_{2R}\nu _{3L}+v_{\sigma }\bar{\nu}%
_{2L}^{c}N_{3R}^{c}+\Lambda _{\sigma }\bar{N}_{2R}N_{3R}^{c}\right.  \notag
\\
&&\left. \,\,\,\,+\lambda _{\sigma }\bar{\nu}_{3L}^{c}\nu _{2L}+v_{\sigma }%
\bar{N}_{3R}\nu _{2L}+v_{\sigma }\bar{\nu}_{3L}^{c}N_{2R}^{c}+\Lambda
_{\sigma }\bar{N}_{3R}N_{2R}^{c}\right]  \notag \\
&+&\frac{1}{2}z\left[ v_{\rho }\bar{\nu}_{2L}^{c}N_{1R}^{c}-v_{\rho }\bar{N}%
_{2R}\nu _{1L}-v_{\rho }\bar{\nu}_{1L}^{c}N_{2R}^{c}+v_{\rho }\bar{N}%
_{1R}\nu _{2L}\right] +H.c.  \label{vienLm}
\end{eqnarray}

We can rewrite (\ref{vienLm}) in the matrix form
\begin{equation}
-\mathcal{L}_{\nu }^{\mathrm{mass}}=\frac{1}{2}\bar{\chi}_{L}^{c}M_{\nu
}\chi _{L}+H.c.,\hspace*{0.5cm}\chi _{L}\equiv \left(
\begin{array}{c}
\nu _{L} \\
N_{R}^{c}%
\end{array}%
\right) ,\hspace*{0.5cm}M_{\nu }\equiv \left(
\begin{array}{cc}
M_{L} & M_{D}^{T} \\
M_{D} & M_{R}%
\end{array}%
\right) ,
\end{equation}%
where $\nu _{L}=(\nu _{1L},\nu _{2L},\nu _{3L})^{T}$, $%
N_{R}=(N_{1R},N_{2R},N_{3R})^{T}$ and
\begin{equation}
M_{L,D,R}=\left(
\begin{array}{ccc}
a_{L,D,R} & c_{L,D,R} & 0 \\
-c_{L,D,R} & 0 & b_{L,D,R} \\
0 & b_{L,D,R} & 0%
\end{array}%
\right) ,
\end{equation}%
with
\begin{eqnarray}
a_{L} &=&\lambda _{\sigma }x,\hspace*{0.5cm}a_{D}=v_{\sigma }x,\hspace*{0.5cm%
}a_{R}=\Lambda _{\sigma }x,  \notag \\
b_{L} &=&\lambda _{\sigma }y,\hspace*{0.5cm}b_{D}=v_{\sigma }y,\hspace*{0.5cm%
}b_{R}=\Lambda _{\sigma }y,  \notag \\
c_{L} &=&0,\hspace*{0.5cm}c_{D}=v_{\rho }z,\hspace*{0.5cm}c_{R}=0.
\label{abcLDR}
\end{eqnarray}%
%

%%%%%%%%%%%%%%%%%%%%%%%%%%%%%%%%%%%%%%%%%%%%

\begin{figure}[h]
\begin{center}
\includegraphics[width=7.0cm, height=4.50cm]{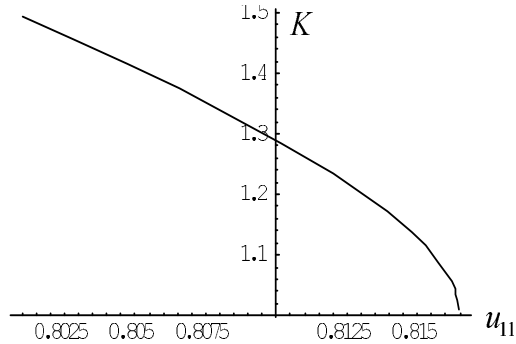}
\vspace*{-0.4cm} \caption[$K$ as a function of
 $u_{11}$ with $u_{1}\in (0.801, 0.845)$ \cite{GonzalezGarcia:2014sz,Conrad:2013dma, Parke:2013pna}]{$K$ as a function of
 $u_{11}$ with $u_{11}\in (0.801, 0.845)$ \cite{GonzalezGarcia:2014sz,Conrad:2013dma, Parke:2013pna}.}\label{KDelta27}
\vspace*{-0.3cm}
\end{center}
\end{figure}

The effective neutrino mass matrix, in the framework of type I and type II seesaw mechanisms,
is given by\footnote{%
With $a_{D,R},b_{D,R}$ given in Eq. (\ref{abcLDR}), $\frac{b_{D}}{b_{R}}-\frac{a_{D}}{a_{R}}=0$, and $({M}_{\mathrm{eff}})_{12}=({M}_{\mathrm{eff}%
})_{21}=\left( \frac{b_{D}}{b_{R}}-\frac{a_{D}}{a_{R}}\right) c_{D}=0$.}
\begin{eqnarray}
{M}_{\mathrm{eff}} &=&M_{L}-M_{D}^{T}{M}_{R}^{-1}M_{D}  \notag \\
&=&\left(
\begin{array}{ccc}
a_{L}-\frac{a_{D}^{2}}{a_{R}} & \hspace*{0.5cm}0 & 0 \\
0 & -\frac{c_{D}^{2}}{a_{R}} & b_{L}-\frac{b_{D}^{2}}{b_{R}} \\
0 & b_{L}-\frac{b_{D}^{2}}{b_{R}} & 0%
\end{array}%
\right) \equiv \left(
\begin{array}{ccc}
A & \hspace*{0.5cm}0 & 0 \\
0 & C & B \\
0 & B & 0%
\end{array}%
\right) ,  \label{Meff}
\end{eqnarray}%
where
\begin{equation*}
A=a_{L}-\frac{a_{D}^{2}}{a_{R}},\hspace*{0.5cm}B=b_{L}-\frac{b_{D}^{2}}{b_{R}%
},\hspace*{0.5cm}C=-\frac{c_{D}^{2}}{a_{R}}.
\end{equation*}%
In the case without the $\rho $ contribution ($v_{\rho }=0$)  we have
$c_{D}=0$ and $M_{\mathrm{eff}}$ in (\ref{Meff}) becomes
\begin{equation}
M_{\mathrm{eff}}^{0}=\left(
\begin{array}{ccc}
A & \hspace*{0.5cm}0 & 0 \\
0 & 0 & B \\
0 & B & 0%
\end{array}%
\right) .  \label{M0nu}
\end{equation}%
The mass matrix in Eq. (\ref{M0nu}) gives the degenerate mass of neutrinos
\begin{equation*}
m_{1}^{0}=-m_{3}^{0}=B,\,\,\,m_{2}^{0}=A,
\end{equation*}%
and the corresponding leptonic mixing matrix yields the tri-bimaximal
mixing form $U_{L}^{+}U_{\nu }=U_{HPS}$, which is ruled out by the recent
neutrino experimental data. However, the $\rho $ contribution will improve
this. Indeed, the mass matrix (\ref{Meff}) is diagonalized as follows $%
U_{\nu }^{T}M_{\mathrm{eff}}U_{\nu }=\mathrm{diag}(m_{1},m_{2},m_{3})$, with
\begin{equation}
m_{1,3}=\frac{1}{2}\left( C\pm \sqrt{C^{2}+4B^{2}}\right) ,\hspace*{0.5cm}%
m_{2}=A,  \label{m123}
\end{equation}%
and the corresponding neutrino mixing matrix:
\begin{equation}
U_{\nu }=\left(
\begin{array}{ccc}
0 & 1 & 0 \\
\frac{K}{\sqrt{K^{2}+1}} & 0 & -\frac{1}{\sqrt{K^{2}+1}} \\
\frac{1}{\sqrt{K^{2}+1}} & 0 & \frac{K}{\sqrt{K^{2}+1}}%
\end{array}%
\right) \left(
\begin{array}{ccc}
1 & 0 & 0 \\
0 & 1 & 0 \\
0 & 0 & i%
\end{array}%
\right) ,  \label{Unu1}
\end{equation}%
where
\begin{equation}
K=\frac{C-\sqrt{C^{2}+4B^{2}}}{2B}.  \label{K}
\end{equation}%
Combining (\ref{Uclepp}) and (\ref{Unu1}), the lepton mixing matrix
takes the form:
\begin{equation}
U_{Lep}=U_{L}^{\prime +}U_{\nu }=\left(
\begin{array}{ccc}
u_{11} & u_{12} & u_{13} \\
u_{21} & u_{22} & u_{23} \\
u_{31} & u_{32} & u_{33}%
\end{array}%
\right) \left(
\begin{array}{ccc}
1 & 0 & 0 \\
0 & 1 & 0 \\
0 & 0 & i%
\end{array}%
\right) ,\label{Ulepm}
\end{equation}%
where
\begin{eqnarray}
u_{11} &=&\frac{K\beta _{2}+\beta _{3}}{\sqrt{K^{2}+1}},\hspace*{0.5cm}%
u_{12}=u_{22}=u_{32}=\beta _{1},  \notag \\
u_{13} &=&\frac{-\beta _{2}+K\beta _{3}}{\sqrt{K^{2}+1}},\hspace*{0.5cm}%
u_{21}=\frac{\omega (K\beta _{2}+\omega \beta _{3})}{\sqrt{K^{2}+1}},%
\hspace*{0.5cm}u_{23}=\frac{\omega (-\beta _{2}+K\omega \beta _{3})}{\sqrt{%
K^{2}+1}},  \notag \\
u_{31} &=&\frac{\omega (K\omega \beta _{2}+\beta _{3})}{\sqrt{K^{2}+1}},%
\hspace*{0.5cm}u_{33}=\frac{\omega (-\omega \beta _{2}+K\beta _{3})}{\sqrt{%
K^{2}+1}},  \label{Ulij}
\end{eqnarray}%
with
\begin{equation*}
\beta _{i}=\frac{1}{3\alpha _{i}}\,\,(i=1,2,3).
\end{equation*}%
We see that all the elements of the matrix $U_{Lep}$ in Eq. (\ref{Ulij}) depend only on two
parameters $\varepsilon _{3}$ ans $K$. From
experimental constraints on the elements of the lepton mixing matrix given
in Refs. \cite{GonzalezGarcia:2014sz,Conrad:2013dma, Parke:2013pna}, we can find out the regions
of $K$ and $\varepsilon _{3}$ that satisfy experimental data on lepton
mixing matrix. Indeed, in the case $\alpha_{i}=\beta_{i}=1/\sqrt{3} \,(i=1,2,3)$ and $K=1$, the
lepton mixing matrix in Eq. (\ref{Ulepm}) reduces to Tri-bimaximal form. Therefore, the realistic lepton mixing pattern can be obtained if the values of $\alpha_{i}, \, \beta_{i} \, (i=1,2,3)$ are close to  $1/\sqrt{3}$ and $K$ gets values close to unity. If $\alpha_{i}=\beta_{i}=1/\sqrt{3} \,(i=1,2,3)$, the element $u_{11}$ in Eq. (\ref{Ulij}) becomes, $u_{11}=\frac{K+1}{\sqrt{3(K^{2}+1)}}$. By using the experimental constraint values of $u_{11}$ given in  \cite%
{GonzalezGarcia:2014sz,Conrad:2013dma, Parke:2013pna}, $0.801 \leq\left|u_{11}\right| \leq 0.845$ we get $1.1 \leq\left|K \right| \leq 1.5$ which is depicted in Fig. \ref{KDelta27}.

To get the specific value of $\varepsilon _{3}$, a specific value of $K$ would be chosen with an experimental value of $u_{11}$. In the case  $K=\sqrt{2}\simeq 1.4142$, combining
with the constraint values on the
element $u_{11}$ of lepton mixing matrix \cite%
{GonzalezGarcia:2014sz,Conrad:2013dma, Parke:2013pna}, $u_{11}=0.805$, we
obtain a solution\footnote{In this model, the choice of the parameters is not unique. It is
just one specific example to show that there exist the model parameters consistent with the experimental data.}:
\begin{equation}
\varepsilon _{3}=-0.000743889+0.000785038i.  \label{ep31}
\end{equation}%
Then, it follows that the Pontecorvo-Maki-Nakagawa-Sakata (PMNS) leptonic mixing matrix takes the form:
\begin{equation}
U_{Lep}\simeq \left(
\begin{array}{ccc}
0.805 & \hspace*{0.5cm}0.577 & \hspace*{0.5cm}0.137988i \\
-0.402851+0.119716i & \hspace*{0.5cm}0.577 & \hspace*{0.5cm}%
0.696899-0.0691182i \\
-0.402149-0.119716i & \hspace*{0.5cm}0.577 & \hspace*{0.5cm}%
-0.697328-0.0688701i%
\end{array}%
\right) .  \label{Ulep}
\end{equation}
which implies that
\begin{equation}
|U_{Lep}|=\left(
\begin{array}{ccc}
0.805 & \hspace*{0.5cm}0.577 & \hspace*{0.5cm}0.137988 \\
0.420263 & \hspace*{0.5cm}0.577 & \hspace*{0.5cm}0.70031 \\
0.41959 & \hspace*{0.5cm}0.577 & \hspace*{0.5cm}0.70072%
\end{array}%
\right).\label{Ulepa1}
\end{equation}%
Using Eq. (\ref{K}) and $K=\sqrt{2}$, we obtain
\begin{equation}
C=\frac{B}{\sqrt{2}}.  \label{BCv1}
\end{equation}
In the standard Particle Data Group(PDG) parametrization, the lepton mixing
matrix can be parametrized as
\begin{equation}
U_{PMNS}=%
\begin{pmatrix}
c_{12}c_{13} & s_{12}c_{13} & s_{13}e^{-i\delta } \\
-s_{12}c_{23}-c_{12}s_{23}s_{13}e^{i\delta } &
c_{12}c_{23}-s_{12}s_{23}s_{13}e^{i\delta } & s_{23}c_{13} \\
s_{12}s_{23}-c_{12}c_{23}s_{13}e^{i\delta } &
-c_{12}s_{23}-s_{12}c_{23}s_{13}e^{i\delta } & \,\,c_{23}c_{13}%
\end{pmatrix}%
\times \mathcal{P},  \label{Ulepg}
\end{equation}%
where $\mathcal{P}=\mathrm{diag}(1,e^{i\alpha },e^{i\beta })$, and $%
c_{ij}=\cos \theta _{ij}$, $s_{ij}=\sin \theta _{ij}$ with $\theta _{12}$, $%
\theta _{23}$ and $\theta _{13}$ being the solar, atmospheric and reactor
angles, respectively. $\delta =[0,2\pi ]$ is the Dirac CP violation phase
while $\alpha $ and $\beta $ are two Majorana CP violation phases. The
observable angles in the standard PMNS parametrization are given by \cite%
{Agashe:2014kda}
\begin{equation}
s_{13}=\left\vert U_{13}\right\vert ,\,\,s_{23}=\frac{\left\vert
U_{23}\right\vert }{\sqrt{1-\left\vert U_{13}\right\vert ^{2}}},\,\,s_{12}=%
\frac{\left\vert U_{12}\right\vert }{\sqrt{1-\left\vert U_{13}\right\vert
^{2}}}.  \label{thetaij}
\end{equation}%
Combining Eqs.(\ref{Ulepa1}) and (\ref{thetaij}) yields:
\begin{equation}
\sin \theta _{13}=0.137988,\,\,\sin \theta _{23}=0.713911,\,\,\sin \theta
_{12}=0.588205,
\end{equation}%
or
\begin{equation}
\theta _{13}\simeq 7.9315^{\circ },\,\,\theta _{23}\simeq 45.5541^{\circ
},\,\,\theta _{12}\simeq 36.0293^{\circ },
\end{equation}%
which are all very consistent with the recent data on neutrino mixing
angles. Furthermore, comparing the lepton mixing matrix given in Eq. (\ref{Ulep}) with the
standard parametrization in Eq.(\ref{Ulepg}), one obtains
vanishing Majorana phases, i.e., $\alpha =0,\beta =0$ as well as
nonvanishing leptonic Dirac CP violating phase $\delta =-\frac{\pi }{2}$ and
Jarskog invariant close to $-3.2\times 10^{-2}$. It is worth mentioning that having leptonic mixing parameters consistent with their experimental values, require that the parameter $K$ to be equal or very close to $\sqrt{2}$. The other parameters that determine the leptonic mixing angles are $Re\left(\varepsilon_3\right)$ and $Im\left(\varepsilon_3\right)$, i.e, which are of the order of $10^{-4}$. Besides that we have numerically checked the leptonic mixing parameters have a low sensitivity with $Re\left(\varepsilon_3\right)$ and $Im\left(\varepsilon_3\right)$ but are highly sensitive under small variations around $K=\sqrt{2}$, for example having $K=0.9\sqrt{2}\simeq 1.27$ leads to $\sin^2\theta_{13}=0.009$, which is outside the $3\sigma$ experimentally allowed range. In the region of parameter space consistent with the experimental values of the leptonic mixing parameters, we have numerically checked that the leptonic Dirac CP violating phase is equal to $-\frac{\pi}{2}$. Other phases different than $-\frac{\pi}{2}$ are obtained for values of the $K$ parameters outside the vicinity of $K=\sqrt{2}$, that leads to a reactor mixing angle $\theta_{13}$ unacceptably small.

At present, the absolute neutrino masses as well as the mass ordering of
neutrinos is unknown. The result in \cite{Tegmark:2003ud} shows that
\begin{equation}
m_i\leq 0.6\, \mathrm{eV},\hspace{1cm}i=1,2,3,  \label{upb}
\end{equation}
while the upper bound on the sum of light active neutrino masses is given by \cite{Weiler:2013rta}
\begin{equation}
\sum^{3}_{i=1} m_i\leq 0.5\, \mathrm{eV}.  \label{upbsum}
\end{equation}

The neutrino mass spectrum can be described by the normal mass hierarchy ($|m_{1}|\simeq
|m_{2}|<|m_{3}|$), the inverted hierarchy ($|m_{3}|<|m_{1}|\simeq |m_{2}|$)
or the nearly degenerate ($|m_{1}|\simeq |m_{2}|\simeq |m_{3}|$) ordering. The neutrino mass
ordering depends on the sign of $\Delta m_{23}^{2}$, which is currently unknown. In the case of
3-neutrino mixing, in the model under consideration, the two possible signs of
$\Delta m_{23}^{2}$ correspond to two types of allowed neutrino mass spectra.

\subsection{Normal case ($\Delta m^2_{23}> 0$)}

Substituting $B$ from (\ref{BCv1}) into (\ref{m123}) and taking
into account the experimental values of the neutrino mass squared splittings for the normal hierarchy given in \cite{Agashe:2014kda}, i.e., $\Delta
m_{21}^{2}=7.53\times 10^{-5}eV^{2},\Delta m_{32}^{2}=2.44\times
10^{-3}eV^{2}$, we get the following solution:
\begin{equation}
A=0.030228,\,\,B=0.0409496,C=0.0289557,  \label{solu1}
\end{equation}%
which implies that:
\begin{equation}
\left\vert m_{1}\right\vert =0.0289557\,\mathrm{eV},\,\,m_{2}=0.030228\,%
\mathrm{eV},\,\,m_{3}=0.0579114\,\mathrm{eV}.  \label{mnuNH}
\end{equation}%
\begin{equation}
x=\frac{0.030228\Lambda _{\sigma }}{\Lambda _{\sigma }\lambda _{\sigma
}-v_{\sigma }^{2}},\hspace*{0.5cm}y=\frac{0.0409496\Lambda _{\sigma }}{%
\Lambda _{\sigma }\lambda _{\sigma }-v_{\sigma }^{2}},\hspace*{0.5cm}z=\frac{%
0.029585i\Lambda _{\sigma }}{v_{\rho }\sqrt{\Lambda _{\sigma }\lambda
_{\sigma }-v_{\sigma }^{2}}}.  \label{vNH}
\end{equation}

\subsection{Inverted case ($\Delta m^2_{23}<0$)}

Substituting $B$ from (\ref{BCv1}) into (\ref{m123}) and taking
into account the neutrino oscillation experimental data of neutrino mass squared differences for the
inverted neutrino mass orderings given in \cite{Agashe:2014kda}, i.e., $\Delta
m_{21}^{2}=7.53\times 10^{-5}eV^{2},\Delta m_{32}^{2}=2.52\times
10^{-3}eV^{2}$, we find the solution:
\begin{equation}
A=0.0577486,\,\,B=-0.0403708,C=-0.0285465,  \label{solu2}
\end{equation}%
which implies that:
\begin{equation}
\left\vert m_{1}\right\vert =0.0570929\,\mathrm{eV},\,\,m_{2}=0.0577486\,%
\mathrm{eV},\,\,m_{3}=0.0285465\,\mathrm{eV}.  \label{mnuIH}
\end{equation}%
\begin{equation}
x=\frac{0.0577486\Lambda _{\sigma }}{\Lambda _{\sigma }\lambda _{\sigma
}-v_{\sigma }^{2}},\hspace*{0.5cm}y=\frac{0.0403708\Lambda _{\sigma }}{%
\Lambda _{\sigma }\lambda _{\sigma }-v_{\sigma }^{2}},\hspace*{0.5cm}z=\frac{%
0.0577486i\Lambda _{\sigma }}{v_{\rho }\sqrt{\Lambda _{\sigma }\lambda
_{\sigma }-v_{\sigma }^{2}}}.  \label{vIH}
\end{equation}

\subsection{Effective Majorana neutrino mass parameter}
In what follows we proceed to compute the effective Majorana neutrino mass
parameter, whose value is proportional to the amplitude of neutrinoless double beta ($0\nu
\beta \beta $) decay. The effective Majorana neutrino mass
parameter has the form:
\begin{equation}
m_{\beta \beta }=\left\vert \sum_{j}U_{ek}^{2}m_{\nu _{k}}\right\vert ,
\label{mee}
\end{equation}%
where $U_{ej}^{2}$ is the squared of the PMNS leptonic mixing matrix elements
 andand $m_{\nu _{k}}$ correspond to the masses of the Majorana neutrinos.

From Eqs. (\ref{mnuNH}), (\ref{mnuIH}), (\ref{Ulep}) and (\ref{mee}), it
follows that the effective Majorana neutrino mass parameter, for the Normal
and Inverted neutrino mass orderings, acquires the following values:
\begin{equation}
m_{\beta \beta }=\left\{
\begin{array}{l}
10\ \mbox{meV}\ \ \ \ \ \ \ \mbox{for \ \ \ \ Normal Hierarchy} \\
18\ \mbox{meV}\ \ \ \ \ \ \ \mbox{for \ \ \ \ Inverted Hierarchy}%
\end{array}%
\right.   \label{meevalues}
\end{equation}%
As seen from Eq. (\ref{meevalues}), the resulting effective Majorana
neutrino mass parameters for normal and inverted neutrino mass orderings, are out the scope of the present and future
$0\nu \beta \beta $ decay experiments. Let us note that the Majorana
neutrino mass parameter has the upper limit $m_{\beta \beta }\leq 160$ meV,
corresponding to $T_{1/2}^{0\nu \beta \beta }(^{136}\mathrm{Xe})\geq
1.6\times 10^{25}$ yr at 90\% C.L, as follows from the EXO-200 experiment
\cite{Auger:2012ar}. That limit is expected to be updated in a not too
distant future. The GERDA \textquotedblleft phase-II\textquotedblright
experiment \cite{Abt:2004yk,Ackermann:2012xja}
is expected to reach
\mbox{$T^{0\nu\beta\beta}_{1/2}(^{76}{\rm Ge}) \geq
2\times 10^{26}$ yr}, corresponding to $m_{\beta \beta }\leq 100$ meV. A
bolometric CUORE experiment, using ${}^{130}Te$ \cite{Alessandria:2011rc},
is currently under construction and its estimated sensitivity is about $%
T_{1/2}^{0\nu \beta \beta }(^{130}\mathrm{Te})\sim 10^{26}$ yr,
corresponding to \mbox{$m_{\beta\beta}\leq 50$ meV.} Besides that, there are
plans for ton-scale next-to-next generation $0\nu \beta \beta $
experiments with $^{136}$Xe \cite{KamLANDZen:2012aa,Albert:2014fya} and $%
^{76}$Ge \cite{Abt:2004yk,Guiseppe:2011me} asserting sensitivities over $%
T_{1/2}^{0\nu \beta \beta }\sim 10^{27}$ yr, corresponding to $m_{\beta
\beta }\sim 12-30$ meV. A review on the theory and phenomenology of neutrinoless double-beta decay can be found in  Ref. \cite%
{Bilenky:2014uka}. It is worth mentioning that our model predicts $T_{1/2}^{0\nu \beta
\beta }$ at the level of sensitivities of the next generation or
next-to-next generation $0\nu \beta \beta $ experiments.

\section{Quark masses\label{quark}}

The $[\mathrm{SU}(3)_{L},\mathrm{U}(1)_{X},\mathrm{U}%
(1)_{\mathcal{L}},\underline{\Delta }(27)]$ assignments for the quark sector of the
model are given in Tab. \ref{Fercon}.
Thus, in order to generate quark masses, we additionally
introduce four extra $\mathrm{SU}(3)_{L}$ scalar triplets, assigned as a $\Delta \left( 27\right)$ anti-triplet $(\eta)$ and a $\Delta \left( 27\right)$ non
trivial singlet  $(\chi)$. The scalar fields participating in
the quark Yukawa interactions:
\begin{eqnarray}
\eta  &=&\left( \eta _{1},\eta _{2},\eta _{3}\right),% \sim \left[ 3,-1/3,-1/3,%\underline{\bar{3}}\right]
\hspace*{0.5cm}\eta _{i}=\left(\eta _{i1}^{0} \,\,,
\eta _{i2}^{-} \,\,, \eta _{i3}^{0}\right)^T, \hspace*{0.5cm}i=1,2,3,\\
\chi  &=&\left(\chi _{1}^{0} \,\,,
\chi _{2}^{-} \,\,,
\chi _{3}^{0}\right)^T,
\end{eqnarray}
where their $G$ assignments are reported in table \ref{Fercon} and the VEV pattern of the $\Delta \left( 27\right) $ triplet $\eta $ is given as
\begin{equation}
\langle \eta \rangle =(\langle \eta _{1}\rangle ,\langle \eta _{2}\rangle
,\langle \eta _{3}\rangle )^T,  \label{etaVEV}
\end{equation}%
with
\begin{equation*}
\langle \eta _{i}\rangle =\left(
\begin{array}{c}
u_{i} \\
0 \\
0%
\end{array}%
\right) ,\,\,(i=1,2,3),\hspace*{0.5cm}\langle \chi \rangle =\left(
\begin{array}{c}
0 \\
0 \\
v_{\chi }%
\end{array}%
\right) .
\end{equation*}
The quark Yukawa interactions are
\begin{eqnarray}
-\mathcal{L}_{q} &=&h_{3}^{d}\bar{Q}_{3L}(\phi d_{R})_{1_{2}}+h_{1}^{u}\bar{Q%
}_{1L}(\phi ^{\ast }u_{R})_{1_{1}}+h_{2}^{u}\bar{Q}_{2L}(\phi ^{\ast
}u_{R})_{1_{3}}  \notag \\
&+&h_{3}^{u}\bar{Q}_{3L}(\eta u_{R})_{1_{2}}+h_{1}^{d}\bar{Q}_{1L}(\eta
^{\ast }d_{R})_{1_{1}}+h_{2}^{d}\bar{Q}_{2L}(\eta ^{\ast }d_{R})_{1_{3}}
\notag \\
&+&f_{3}\bar{Q}_{3L}\chi U_{R}+f_{1}\bar{Q}_{1L}\chi ^{\ast }D_{1R}+f_{2}%
\bar{Q}_{2L}\chi ^{\ast }D_{2R}+H.c . \label{quarkintract}
\end{eqnarray}
%%%%%%%%%%%%%%%%%%%%%%%%%%%%%%%%%%%%%%
Then, it follows that the quark mass terms take the form
\begin{eqnarray}
-\mathcal{L}_{q}^{mass} &=&-h_{1}^{u}v_{1}^{\ast }\bar{u}%
_{1L}u_{1R}-h_{1}^{u}v_{2}^{\ast }\bar{u}_{1L}u_{2R}-h_{1}^{u}v_{3}^{\ast }%
\bar{u}_{1L}u_{3R}  \notag \\
&-&h_{2}^{u}v_{1}^{\ast }\bar{u}_{2L}u_{1R}-\omega ^{2}h_{2}^{u}v_{2}^{\ast }%
\bar{u}_{2L}u_{2R}-\omega h_{2}^{u}v_{3}^{\ast }\bar{u}_{2L}u_{3R}  \notag \\
&+&h_{3}^{u}u_{1}\bar{u}_{3L}u_{1R}+\omega h_{3}^{u}u_{2}\bar{u}%
_{3L}u_{2R}+\omega ^{2}h_{3}^{u}u_{3}\bar{u}_{3L}u_{3R}  \notag \\
&+&h_{1}^{d}u_{1}^{\ast }\bar{d}_{1L}d_{1R}+h_{1}^{d}u_{2}^{\ast }\bar{d}%
_{1L}d_{2R}+h_{1}^{d}u_{3}^{\ast }\bar{d}_{1L}d_{3R}  \notag \\
&+&h_{2}^{d}u_{1}^{\ast }\bar{d}_{2L}d_{1R}+\omega ^{2}h_{2}^{d}u_{2}^{\ast }%
\bar{d}_{2L}d_{2R}+\omega h_{2}^{d}u_{3}^{\ast }\bar{d}_{2L}d_{3R}  \notag \\
&+&h_{3}^{d}v_{1}\bar{d}_{3L}d_{1R}+\omega h_{3}^{d}v_{2}\bar{d}%
_{3L}d_{2R}+\omega ^{2}h_{3}^{d}v_{3}\bar{d}_{3L}d_{3R}  \notag \\
&+&f_{3}v_{\chi }\bar{U}_{L}U_{R}+f_{1}v_{\chi }^{\ast }\bar{D}%
_{1L}D_{1R}+f_{2}v_{\chi }^{\ast }\bar{D}_{2L}D_{2R}+H.c.
\end{eqnarray}%
Consequently, the exotic quarks do not mix with the SM quarks. From the quark mass
terms given above, it follows that the exotic quark masses are
\begin{equation*}
m_{U}=|f_{3}v_{\chi }|,\hspace*{0.5cm}m_{D_{1,2}}=|f_{1,2}v_{\chi }^{\ast }|.
\end{equation*}%
and the SM up-type and down-type quark mass matrices take the form:
\begin{equation}
M_{u}=\left(
\begin{array}{ccc}
-h_{1}^{u}v_{1}^{\ast } & \,\,\,-h_{1}^{u}v_{2}^{\ast } &
-h_{1}^{u}v_{3}^{\ast } \\
-h_{2}^{u}v_{1}^{\ast } & \,\,-h_{2}^{u}v_{2}^{\ast }\omega ^{2} &
\,-h_{2}^{u}v_{3}^{\ast }\omega  \\
\,\,\,\,h_{3}^{u}u_{1} & \,\,\,h_{3}^{u}u_{2}\omega  & h_{3}^{u}u_{3}\omega
^{2}%
\end{array}%
\right) ,\,\,\,M_{d}=\left(
\begin{array}{ccc}
h_{1}^{d}u_{1}^{\ast } & \,\,\,\,\,h_{1}^{d}u_{2}^{\ast } &
\,\,\,h_{1}^{d}u_{3}^{\ast } \\
h_{2}^{d}u_{1}^{\ast } & \,\,\,\,h_{2}^{d}u_{2}^{\ast }\omega ^{2} &
\,\,\,\,\,h_{2}^{d}u_{3}^{\ast }\omega  \\
h_{3}^{d}v_{1} & h_{3}^{d}v_{2}\omega  & h_{3}^{d}v_{3}\omega ^{2}%
\end{array}%
\right) .  \label{MuMd}
\end{equation}%
In the quark sector, we assume that the $\Delta (27)$ discrete group is broken down to
the $Z_{3}$ subgroup, which consists of the elements \{$1,b,b^{2}$\}. This breaking is
triggered by the $\Delta(27)$ scalar triplet $\eta$, with the VEV alignment described in Eq. (\ref{etaVEV}). %
In the case $v_{1}=v_{2}=v_{3},\,u_{1}=u_{2}=u_{3}$ and $v_{i}^{\ast
}=v_{i},\,u_{i}^{\ast }=u_{i}\,\,(i=1,2,3)$, the matrices $M_{u}$ and $M_{d}$ given by Eq. (\ref{MuMd})
are diagonalized by the unitary matrices
\begin{equation*}
V_{0R}^{u}=V_{0R}^{d}=\frac{1}{3}\left(
\begin{array}{ccc}
1 & 1 & 1 \\
1 & \omega  & \omega ^{2} \\
1 & \omega ^{2} & \omega
\end{array}%
\right) ,\hspace*{0.5cm}V_{L}^{u}=V_{L}^{d}=1,
\end{equation*}%
and the quark mixing matrix $V_{\mathrm{CKM}}=V_{L}^{d\dagger
}V_{L}^{u}=1$, which is acceptable since the quark mixing matrix is
very close to the identity matrix \cite{Agashe:2014kda}. By an appropriate
choice of parameters in the SM quark mass matrices given by Eq. (\ref{MuMd}%
), we can successfully reproduce the experimental values of quark masses and
quark mixing angles.
Furthermore it is noteworthy to mention that our model is an extension of the 3-3-1 model considered in \cite{Okada:2016whh}. As pointed out in Refs. \cite{Okada:2016whh}, the flavor constraints can be fullfilled by considering the scale of breaking of the $SU(3)_{L}\otimes U(1)_{X}$ gauge symmetry much larger than the electroweak symmetry breaking scale $v=246$ GeV, which corresponds to the alignment limit of the mass matrix for the CP-even Higgs bosons. Consequently, following \cite{Okada:2016whh}, we expect that the FCNC effects as well as the constraints arising from $K^0-\bar{K^0}$, $B^0-\bar{B^0}$ and $D^0-\bar{D^0}$ mixings will be fullfilled in our model, by considering the scale of breaking of the $SU(3)_{L}\otimes U(1)_{X}$ gauge symmetry much larger than scale of breaking of the electroweak symmetry. In that alignment limit, our model effectively becomes a nine Higgs doublet model, whose scalar sector includes 9 CP even neutral Higges, 8 CP odd neutral Higges and 16 charged Higges. That scalar sector is not predictive as its corresponding scalar potential has many free uncorrelated parameters that can be adjusted to get the required pattern of scalar masses. Therefore, the loop effects of the heavy scalars contributing to certain observables can be suppressed by the appropriate choice of the free parameters in the scalar potential. Fortunately, all these adjustments do not affect the charged fermion and neutrino sector, which is completely controlled by the fermion-Higgs Yukawa couplings. In addition, in models with discrete flavor symmetries, like ours, the deviation of the CKM matrix from the identity can be given by the FCNC effects with the left-handed quarks, but in the alignment limit previously described, such deviations are highly suppressed by the mass of the extra quarks \cite{Dong:2010zu}.

\newpage
\section{Conclusions\label{conclusion}}

\appendix We constructed the first $SU(3)_{C}\otimes SU(3)_{L}\otimes
U(1)_{X}$ model based on the $\Delta \left( 27\right) $ flavor symmetry
supplemented by the $\mathrm{U}(1)_{\mathcal{L}}$ new lepton global
symmetry. This $\mathrm{U}(1)_{\mathcal{L}}$ new lepton global symmetry
allows us to have different scalar fields in the Yukawa interactions for
charged lepton, neutrino and quark sectors, thus allowing us to treat these
sectors independently. Our model successfully accounts for fermion masses
and mixings. In our model, the neutrino Yukawa interactions include three $%
SU(3)_{L}$ scalar triplets as well as three $SU(3)_{L}$ scalar antisextets
that allow to implement type II and type I seesaw mechanisms, respectively,
for the generation of the light active neutrino masses. Consequently, light
active neutrino masses arise from a combination of type-I and type-II seesaw
mechanisms, mediated by three heavy right handed Majorana neutrinos and
three $SU(3)_{L}$ scalar antisextets, respectively. Furthermore, from the consistency of leptonic mixing angles with their experimental values, we obtain a non vanishing leptonic Dirac CP violating phase equal to $-\frac{\pi }{2}$. In addition, our model features an effective Majorana neutrino mass parameter of neutrinoless double beta decay, with values $m_{\beta \beta }=$ 10 and 18 meV for the normal and the inverted neutrino mass hierarchies, respectively.

\section*{Acknowledgments}
This research has received funding from the Vietnam National Foundation for Science and Technology Development (NAFOSTED) under grant number 103.01-2015.33. A.E.C.H was supported by DGIP internal Grant No. 111458 and by Proyecto Basal FB0821.  H. N. Long thanks Universidad T\'{e}cnica Federico Santa Mar\'{\i}a for hospitality, where this work was finished. The visit of H. N.  Long to Universidad T\'{e}cnica Federico Santa Mar\'{\i}a was supported by DGIP internal Grant No. 111458.

\newpage
\section{$\Delta (27)$ group and Clebsch-Gordan coefficients\label{Delta27g}}

The $\Delta (27)$ discrete group is a subgroup of $SU(3)$ and is isomorphic
to the semi-direct product group $(Z^{\prime}_3\times Z^{\prime
\prime}_3)\rtimes Z_3$. It is also a simple group \footnote{%
In fact, the simplest group of the type $\Delta (3n^2)$ is $\Delta (3)\equiv
Z_3$. The next group, $\Delta(12)$, is isomorphic to $A_4$. Thus, the
simplest non-trivial group of the type $\Delta (3n^2)$ is $\Delta(27)$.} of
the type $\Delta (3n^2)$ with $n=3$. The $\Delta (27)$ discrete group has 27 elements divided
into 11 conjugacy classes, so it has 11 irreducible representations,
including two triplets ($\underline{3}$ and its conjugate $\underline{\bar{3}%
}$) and 9 singlets $\underline{1}_i \,\, (i=1,2,..., 9)$. Any element of $%
\Delta (27)$ can be written as a multiplication of three generators, i.e., $b, a$ and
$a^{\prime }$, in the form $b^k a^m {a^{\prime}}^n$, satisfying the
relations
\begin{eqnarray}
&& a^3 =a^{\prime 3}=b^3=1,\hspace*{0.5cm} a a^{\prime }=a^{\prime }a,
\notag \\
&& bab^{-1}=(a^{\prime }a)^{-1},\,\, b a^{\prime}b^{-1}=a,
\label{aapbrela}
\end{eqnarray}
where $b$ is a generator of $Z_3$, and $a$, $a^{\prime }$ belong to $%
Z^{\prime}_3$ and $Z^{\prime \prime}_3$, respectively.

The character table of $\Delta (27)$ is given in Tab. \ref{CTD27}, where $n$
is the number of elements, $h$ is the order of each element, and $\omega =e^{%
\frac{2\pi i}{3}}=-\frac{1}{2}+i\frac{\sqrt{3}}{2}$ is the cube root of
unity, obeying $1+\omega+\omega^2=0$ and $\omega^3=1$.
\begin{table}[ht]
\caption{Character table of $\Delta(27)$.}
\label{CTD27}
\begin{center}
\resizebox{15.5cm}{!}{
\begin{tabular}{|c|c|c||c|c|c|c|c|c|c|c|c|c|c|}
\hline
\textbf{Class} & \,\, $n$ \,\, & \, \,\,$h$ \,\, & \, \,\,$\underline{1}_1$
\,\, & \,\,\, $\underline{1}_2$ \,\, & \, \,\,$\underline{1}_3$ \,\, & \,\, $\underline{1}_4$ \,\, & \,\, $\underline{1}_5$ \,\, & \,\,$\underline{1}_6$ \,\, & \,\, $\underline{1}_7$ \,\, & \,\, $\underline{1}_8$ \,\, & \,\, $\underline{1}_9$ \,\, & \,\, $\underline{3}$ \,\, & \,\, $\bar{\underline{3}}$\,\,\, \\ \hline\hline
$C_1$ & 1 & 1 & 1 & 1 & 1 & 1 & 1 & 1 & 1 & 1 & 1 & 3 & 3 \\ \hline
$C_2$ & 1 & 3 & 1 & 1 & 1 & 1 & 1 & 1 & 1 & 1 & 1 & 3$\omega$ & 3$\omega^2$
\\ \hline
$C_3$ & 1 & 3 & 1 & 1 & 1 & 1 & 1 & 1 & 1 & 1 & 1 & 3$\omega^2$ & 3$\omega$
\\ \hline
$C_4$ & 3 & 3 & 1 & $\omega$ & $\omega^2$ & 1 & $\omega^2$ & $\omega$ & 1 & $%
\omega$ & $\omega^2$ & 0 & 0 \\ \hline
$C_5$ & 3 & 3 & 1 & $\omega^2$ & $\omega$ & 1 & $\omega$ & $\omega^2$ & 1 & $%
\omega^2$ & $\omega$ & 0 & 0 \\ \hline
$C_6$ & 3 & 3 & 1 & 1 & 1 & $\omega^2$ & $\omega^2$ & $\omega^2$ & $\omega$
& $\omega$ & $\omega$ & 0 & 0 \\ \hline
$C_7$ & 3 & 3 & 1 & $\omega$ & $\omega^2$ & $\omega^2$ & $\omega$ & 1 & $%
\omega$ & $\omega^2$ & 1 & 0 & 0 \\ \hline
$C_8$ & 3 & 3 & 1 & $\omega^2$ & $\omega$ & $\omega^2$ & 1 & $\omega$ & $%
\omega$ & 1 & $\omega^2$ & 0 & 0 \\ \hline
$C_9$ & 3 & 3 & 1 & 1 & 1 & $\omega$ & $\omega$ & $\omega$ & $\omega^2$ & $%
\omega^2$ & $\omega^2$ & 0 & 0 \\ \hline
$C_{10}$ & 3 & 3 & 1 & $\omega^2$ & $\omega$ & $\omega$ & $\omega^2$ & 1 & $%
\omega^2$ & $\omega$ & 1 & 0 & 0 \\ \hline
$C_{11}$ & 3 & 3 & 1 & $\omega$ & $\omega^2$ & $\omega$ & 1 & $\omega^2$ & $%
\omega^2$ & 1 & $\omega$ & 0 & 0 \\ \hline
\end{tabular}}
\end{center}
\end{table}
The conjugacy classes generated from $b, a$ and $a^{\prime }$ are presented
in Eq. (\ref{CCD27}).

\begin{eqnarray}
C_1 &:& \{e\},\hspace*{0.5cm}\hspace*{0.5cm}\hspace*{0.5cm}\hspace*{0.5cm}%
\hspace*{0.5cm}\hspace*{0.5cm}\hspace*{0.5cm}\,\, h=1,  \notag \\
C_2 &:& \{a^2 a^{\prime }\}, \hspace*{0.5cm}\hspace*{0.5cm}\hspace*{0.5cm}%
\hspace*{0.5cm}\hspace*{0.5cm}\hspace*{0.5cm}\,\, h=3,  \notag \\
C_3 &:& \{a a^{\prime 2}\}, \hspace*{0.5cm}\hspace*{0.5cm}\hspace*{0.5cm}%
\hspace*{0.5cm}\hspace*{0.5cm}\hspace*{0.5cm}\,\,\, h=3,  \notag \\
C_4 &:& \{b,\, ba^2 a^{\prime },\, ba a^{\prime 2}\}, \hspace*{0.5cm}%
\hspace*{0.5cm}\hspace*{0.5cm} h=3,  \notag \\
C_5 &:& \{b^2,\, b^2 a^2 a^{\prime 2 }a a^{\prime 2}\}, \hspace*{0.5cm}%
\hspace*{0.5cm} h=3,  \notag \\
C_6 &:& \{aa^{\prime 2 },\, a^{\prime 2}\}, \hspace*{0.5cm}\hspace*{0.5cm}%
\hspace*{0.5cm}\hspace*{0.5cm}\, h=3,  \notag \\
C_7 &:& \{ba^2,\, ba^{\prime 2},\, b a a^{\prime }\}, \hspace*{0.5cm}%
\hspace*{0.5cm}\hspace*{0.5cm}\, h=3,  \notag \\
C_8 &:& \{b^2a^{\prime 2},\, b^2 a a^{\prime 2 }a^2\}, \hspace*{0.5cm}%
\hspace*{0.5cm}\, h=3,  \notag \\
C_9 &:& \{a^2a^{\prime 2},\, a,\, a^{\prime }\}, \hspace*{0.5cm}\hspace*{%
0.5cm}\hspace*{0.5cm}\hspace*{0.5cm}\, h=3,  \notag \\
C_{10} &:& \{ba,\,ba^{\prime},  \, ba^{-1}a^{\prime 2}\}, \hspace*{0.5cm}\hspace*{%
0.5cm} h=3,  \notag \\
C_{11} &:& \{b^2a^{\prime}, \, b^2a^{-1}a^{\prime 2},\, b^2 a\}, \hspace*{0.5cm}
h=3.  \label{CCD27}
\end{eqnarray}
The multiplication rules for $\Delta(27)$ group are
\begin{eqnarray}
\underline{3}\otimes \underline{3}= \underline{\bar{3}} ( x_{1}y_1 ,\,\,
x_{2}y_2, \,\, x_{3}y_3) &\oplus& \underline{\bar{3}} (x_{2}y_3+x_3y_2 ,\,\,
x_{3}y_1+x_1y_3, \,\, x_{1}y_2+x_2y_1 )  \notag \\
&\oplus& \underline{\bar{3}} \left( x_{2}y_3- x_{3}y_2, \,\,
x_{3}y_1-x_{1}y_3 ,\,\, x_{1}y_2-x_{2}y_1 \right),  \label{33multip}
\end{eqnarray}
and
\begin{eqnarray}
\underline{3}\otimes \underline{\bar{3}}=\sum_{i=1}^9 \oplus\underline{1}_i ,
\end{eqnarray}
where
\begin{eqnarray}
\underline{1}_1&=&x_1\bar{y_1}+x_2\bar{y_2}+x_3\bar{y_3}, ~~~~~~~~~~~~~ \underline{1}_2=x_1\bar{y_1}%
+\omega x_2\bar{y_2}+\omega^2x_3\bar{y_3},  \notag \\
\underline{1}_3 &=&x_1\bar{y_1}+\omega^2x_2\bar{y_2}+\omega x_3\bar{y_3},~~~~~~~~\underline{1}_4=x_1%
\bar{y_2}+x_2\bar{y_3}+x_3\bar{y_1},  \notag \\
\underline{1}_5&=&x_1\bar{y_2}+\omega x_2\bar{y_3}+\omega^2 x_3\bar{y_1}, ~~~~~~~~
\underline{1}_6=x_1\bar{y_2}+\omega^2 x_2\bar{y_3}+\omega x_3\bar{y_1},  \notag \\
\underline{1}_7&=&x_2\bar{y_1}+ x_3\bar{y_2}+ x_1\bar{y_3}, ~~~~~~~~~~~~~ \underline{1}_8=x_2\bar{y_1%
}+\omega^2 x_3\bar{y_2}+\omega x_1\bar{y_3},  \notag \\
\underline{1}_9&=&x_2\bar{y_1}+\omega x_3\bar{y_2}+\omega^2 x_1\bar{y_3},
\label{multiplication2}
\end{eqnarray}
with $\omega =e^{2\pi i/3}\equiv-\frac{1}{2}+i\frac{\sqrt{3}}{2}$. The
singlets multiplications are given in Table \ref{D27multiplets}.
\begin{table}[h]
\begin{center}
\caption{\label{D27multiplets} The singlet multiplications of the group $\Delta(27)$.}
\begin{tabular}{|c||c|c|c|c|c|c|c|c|c|c|c|}
\hline
  \textbf{Singlets} &~ $\underline{1}_2$ ~ & ~ $\underline{1}_3$ ~ & ~ $\underline{1}_4$ ~
              & ~ $\underline{1}_5$ ~ &~ $\underline{1}_6$~ & ~ $\underline{1}_7$ ~ & ~ $\underline{1}_8$ ~ & ~ $\underline{1}_9$~ \\ \hline\hline
~$\underline{1}_2$ ~    &$\underline{1}_3$& $\underline{1}_1$  & $\underline{1}_6$       &$\underline{1}_4$  &$\underline{1}_5$&$\underline{1}_8$ & $\underline{1}_9$& $\underline{1}_7$     \\ \hline
$\underline{1}_3$  &$\underline{1}_1$& $\underline{1}_2$&$\underline{1}_5$&$\underline{1}_6$  &$\underline{1}_4$& $\underline{1}_9$&$\underline{1}_7$ &$\underline{1}_8$      \\ \hline
$\underline{1}_4$     &$\underline{1}_6$&    $\underline{1}_5$      &$\underline{1}_7$   &$\underline{1}_9$ &$\underline{1}_8$&$\underline{1}_1$   &$\underline{1}_2$    &$\underline{1}_3$     \\ \hline
$\underline{1}_5$    &$\underline{1}_4$& $\underline{1}_6$      &$\underline{1}_9$ &$\underline{1}_8$  &$\underline{1}_7$&$\underline{1}_3$&$\underline{1}_1$&$\underline{1}_2$      \\ \hline
$\underline{1}_6$   &$\underline{1}_5$   &$\underline{1}_4$      &$\underline{1}_8$   &$\underline{1}_7$ &$\underline{1}_9$&$\underline{1}_2$&$\underline{1}_3$&$\underline{1}_1$      \\ \hline
$\underline{1}_7$&  $\underline{1}_8$&  $\underline{1}_9$      &$\underline{1}_1$ &$\underline{1}_3$ &$\underline{1}_2$      &$\underline{1}_4$ &$\underline{1}_6$&$\underline{1}_5$   \\ \hline
$\underline{1}_8$&  $\underline{1}_9$&  $\underline{1}_7$      &$\underline{1}_2$ &$\underline{1}_1$ &$\underline{1}_3$      &$\underline{1}_6$ & $\underline{1}_5$&$\underline{1}_4$   \\ \hline
$\underline{1}_9$&  $\underline{1}_7$&  $\underline{1}_8$      &$\underline{1}_3$ &$\underline{1}_2$
              &$\underline{1}_1$      &$\underline{1}_5$ & $\underline{1}_4$    & $\underline{1}_6$ \\ \hline
\end{tabular}
\end{center}
\end{table}
\newline
It is worth mentioning that $\underline{3} \otimes
\underline{3} \otimes \underline{3}$ has three invariants under the $\Delta(27)$ discrete group. Those invariants are $111 +
222 + 333$, $123 + 231 + 312 - 213 - 321 - 132$ and $123 + 231 + 312 + 213 +
321 + 132$. This is a good feature of the $\Delta(27)$ discrete group, that allows us to make invariant Yukawa couplings to generate fermion mass matrices.

\section{ The matrices of the \textbf{3} representation of $\Delta (27)$%
\label{Delta27repre3}}
The matrices of the $\Delta (27)$ triplet representation are given by:
\begin{eqnarray}
C_{1} &:&\left(
\begin{array}{ccc}
1 & 0 & 0 \\
0 & 1 & 0 \\
0 & 0 & 1%
\end{array}%
\right) ,\,C_{2}:\left(
\begin{array}{ccc}
\omega  & 0 & 0 \\
0 & \omega  & 0 \\
0 & 0 & \omega
\end{array}%
\right) ,\,C_{3}:\left(
\begin{array}{ccc}
\omega ^{2} & 0 & 0 \\
0 & \omega ^{2} & 0 \\
0 & 0 & \omega ^{2}%
\end{array}%
\right) ,  \label{C123} \\
C_{4} &:&\left(
\begin{array}{ccc}
0 & 1 & 0 \\
0 & 0 & 1 \\
1 & 0 & 0%
\end{array}%
\right) ,\hspace*{0.5cm}\,\,\,\,\left(
\begin{array}{ccc}
0 & \omega  & 0 \\
0 & 0 & \omega  \\
\omega  & 0 & 0%
\end{array}%
\right) ,\hspace*{0.5cm}\left(
\begin{array}{ccc}
0 & \omega ^{2} & 0 \\
0 & 0 & \omega ^{2} \\
\omega ^{2} & 0 & 0%
\end{array}%
\right) ,  \label{C4} \\
C_{5} &:&\left(
\begin{array}{ccc}
0 & 0 & 1 \\
1 & 0 & 0 \\
0 & 1 & 0%
\end{array}%
\right) ,\hspace*{0.5cm}\,\,\,\,\left(
\begin{array}{ccc}
0 & 0 & \omega  \\
\omega  & 0 & 0 \\
0 & \omega  & 0%
\end{array}%
\right) ,\hspace*{0.5cm}\left(
\begin{array}{ccc}
0 & 0 & \omega ^{2} \\
\omega ^{2} & 0 & 0 \\
0 & \omega ^{2} & 0%
\end{array}%
\right) ,  \label{C5} \\
C_{6} &:&\left(
\begin{array}{ccc}
1 & 0 & 0 \\
0 & \omega  & 0 \\
0 & 0 & \omega ^{2}%
\end{array}%
\right) ,\hspace*{0.5cm}\left(
\begin{array}{ccc}
\omega ^{2} & 0 & 0 \\
0 & 1 & 0 \\
0 & 0 & \omega
\end{array}%
\right) ,\hspace*{0.5cm}\left(
\begin{array}{ccc}
\omega  & 0 & 0 \\
0 & \omega ^{2} & 0 \\
0 & 0 & 1%
\end{array}%
\right) ,  \label{C6} \\
C_{7} &:&\left(
\begin{array}{ccc}
0 & 1 & 0 \\
0 & 0 & \omega  \\
\omega ^{2} & 0 & 0%
\end{array}%
\right) ,\hspace*{0.5cm}\left(
\begin{array}{ccc}
0 & \omega ^{2} & 0 \\
0 & 0 & 1 \\
\omega  & 0 & 0%
\end{array}%
\right) ,\hspace*{0.5cm}\left(
\begin{array}{ccc}
0 & \omega  & 0 \\
0 & 0 & \omega ^{2} \\
1 & 0 & 0%
\end{array}%
\right) ,  \label{C7} \\
C_{8} &:&\left(
\begin{array}{ccc}
0 & 0 & 1 \\
\omega  & 0 & 0 \\
0 & \omega ^{2} & 0%
\end{array}%
\right) ,\hspace*{0.5cm}\left(
\begin{array}{ccc}
0 & 0 & \omega ^{2} \\
1 & 0 & 0 \\
0 & \omega  & 0%
\end{array}%
\right) ,\hspace*{0.5cm}\left(
\begin{array}{ccc}
0 & 0 & \omega  \\
\omega ^{2} & 0 & 0 \\
0 & 1 & 0%
\end{array}%
\right) ,  \label{C8}\\
C_{9} &:&\left(
\begin{array}{ccc}
1 & 0 & 0 \\
0 & \omega ^{2} & 0 \\
0 & 0 & \omega
\end{array}%
\right) ,\hspace*{0.5cm}\left(
\begin{array}{ccc}
\omega  & 0 & 0 \\
0 & 1 & 0 \\
0 & 0 & \omega ^{2}%
\end{array}%
\right) ,\hspace*{0.5cm}\left(
\begin{array}{ccc}
\omega ^{2} & 0 & 0 \\
0 & \omega  & 0 \\
0 & 0 & 1%
\end{array}%
\right) , \\
C_{10} &:&\left(
\begin{array}{ccc}
0 & 1 & 0 \\
0 & 0 & \omega ^{2} \\
\omega  & 0 & 0%
\end{array}%
\right) ,\hspace*{0.5cm}\left(
\begin{array}{ccc}
0 & \omega  & 0 \\
0 & 0 & 1 \\
\omega ^{2} & 0 & 0%
\end{array}%
\right) ,\hspace*{0.5cm}\left(
\begin{array}{ccc}
0 & \omega ^{2} & 0 \\
0 & 0 & \omega  \\
1 & 0 & 0%
\end{array}%
\right) , \\
C_{11} &:&\left(
\begin{array}{ccc}
0 & 0 & 1 \\
\omega ^{2} & 0 & 0 \\
0 & \omega  & 0%
\end{array}%
\right) ,\hspace*{0.5cm}\left(
\begin{array}{ccc}
0 & 0 & \omega  \\
1 & 0 & 0 \\
0 & \omega ^{2} & 0%
\end{array}%
\right) ,\hspace*{0.5cm}\left(
\begin{array}{ccc}
0 & 0 & \omega ^{2} \\
\omega  & 0 & 0 \\
0 & 1 & 0%
\end{array}%
\right) .
\end{eqnarray}
\section{\label{D27breaking3} The breaking patterns of $\Delta(27)$ by triplets}

For $\Delta(27)$ triplets $\underline{3}$ we have the following VEV alignments:

\begin{itemize}
\item[(1)] \textbf{The first alignment}: ($\langle \phi_1\rangle, \langle
\phi_2\rangle, \langle \phi_3\rangle$) then $\Delta(27)$ is broken
into $\{e\}\equiv\{\mathrm{identity}\}$, i.e., it is completely broken.

\item[(2)] \textbf{The second alignment}: ($\langle \phi_1\rangle, \langle
\phi_1\rangle, \langle \phi_1\rangle$) then $\Delta(27)$ is broken
into $Z_3$ group which consisting of the elements \{$1, b, b^2$\}.

\item[(3)] \textbf{The third alignment}: ($\phi_1\rangle, \langle
\phi_2\rangle, \langle\phi_2\rangle$) or ($\langle
\phi_1\rangle, \langle \phi_2\rangle, \langle \phi_1\rangle$) or ($\langle \phi_1\rangle,\langle \phi_1\rangle,\langle\phi_3\rangle$) then $\Delta(27)$ is completely broken.

\item[(4)] \textbf{The fourth alignment: }($\langle \phi_1\rangle, 0, \langle \phi_3\rangle$) or ($0,\langle
\phi_2\rangle, \langle \phi_3\rangle$) or ($\langle \phi_1\rangle, \langle \phi_2\rangle, 0$) then $\Delta(27)$ is completely broken.

\item[(5)] \textbf{The fifth alignment:} ($\langle \phi_1\rangle, 0, \langle \phi_1\rangle$) or ($0, \langle \phi_2\rangle, \langle \phi_2\rangle$) or ($\langle \phi_1\rangle, \langle \phi_1\rangle, 0$) then $\Delta(27)$ is completely broken.

\item[(6)] \textbf{The sixth alignment:} ($\langle \phi_1\rangle, 0, 0$) or ($0, \langle
\phi_2\rangle,0$) or ($0, 0, \langle \phi_3\rangle$) then
 $\Delta(27)$ is broken into $Z_3$ groups, consisting of the elements
\{$e, aa^{\prime }, (aa^{\prime})^2$\} or \{$e, a, a^2$\} or \{$e, a^{\prime
}, a^{\prime 2}$\}, respectively.
\end{itemize}

Let us note that the breakings of $\Delta(27)$ under $\underline{3}$ and $\underline{\bar{3}}$ are the same.


\begin{thebibliography}{9}
%\cite{Aad:2012tfa}
\bibitem{Aad:2012tfa}
  G.~Aad {\it et al.} [ATLAS Collaboration],
  %``Observation of a new particle in the search for the Standard Model Higgs boson with the ATLAS detector at the LHC,''
  Phys.\ Lett.\ B {\bf 716}, 1 (2012)
  doi:10.1016/j.physletb.2012.08.020
  [arXiv:1207.7214 [hep-ex]].
  %%CITATION = doi:10.1016/j.physletb.2012.08.020;%%
  %5406 citations counted in INSPIRE as of 13 Jan 2016



%\cite{Chatrchyan:2012xdj}
\bibitem{Chatrchyan:2012xdj}
  S.~Chatrchyan {\it et al.} [CMS Collaboration],
  %``Observation of a new boson at a mass of 125 GeV with the CMS experiment at the LHC,''
  Phys.\ Lett.\ B {\bf 716}, 30 (2012)
  doi:10.1016/j.physletb.2012.08.021
  [arXiv:1207.7235 [hep-ex]].
  %%CITATION = doi:10.1016/j.physletb.2012.08.021;%%
  %5300 citations counted in INSPIRE as of 13 Jan 2016



%\cite{An:2012eh}
\bibitem{An:2012eh}
  F.~P.~An {\it et al.} [Daya Bay Collaboration],
  %``Observation of electron-antineutrino disappearance at Daya Bay,''
  Phys.\ Rev.\ Lett.\  {\bf 108}, 171803 (2012)
  doi:10.1103/PhysRevLett.108.171803
  [arXiv:1203.1669 [hep-ex]].
  %%CITATION = doi:10.1103/PhysRevLett.108.171803;%%
  %1366 citations counted in INSPIRE as of 13 Jan 2016



%\cite{Abe:2011sj}
\bibitem{Abe:2011sj}
  K.~Abe {\it et al.} [T2K Collaboration],
  %``Indication of Electron Neutrino Appearance from an Accelerator-produced Off-axis Muon Neutrino Beam,''
  Phys.\ Rev.\ Lett.\  {\bf 107}, 041801 (2011)
  doi:10.1103/PhysRevLett.107.041801
  [arXiv:1106.2822 [hep-ex]].
  %%CITATION = doi:10.1103/PhysRevLett.107.041801;%%
  %1085 citations counted in INSPIRE as of 13 Jan 2016



%\cite{Adamson:2011qu}
\bibitem{Adamson:2011qu}
  P.~Adamson {\it et al.} [MINOS Collaboration],
  %``Improved search for muon-neutrino to electron-neutrino oscillations in MINOS,''
  Phys.\ Rev.\ Lett.\  {\bf 107}, 181802 (2011)
  doi:10.1103/PhysRevLett.107.181802
  [arXiv:1108.0015 [hep-ex]].
  %%CITATION = doi:10.1103/PhysRevLett.107.181802;%%
  %600 citations counted in INSPIRE as of 13 Jan 2016



%\cite{Abe:2011fz}
\bibitem{Abe:2011fz}
  Y.~Abe {\it et al.} [Double Chooz Collaboration],
  %``Indication for the disappearance of reactor electron antineutrinos in the Double Chooz experiment,''
  Phys.\ Rev.\ Lett.\  {\bf 108}, 131801 (2012)
  doi:10.1103/PhysRevLett.108.131801
  [arXiv:1112.6353 [hep-ex]].
  %%CITATION = doi:10.1103/PhysRevLett.108.131801;%%
  %774 citations counted in INSPIRE as of 13 Jan 2016



%\cite{Ahn:2012nd}
\bibitem{Ahn:2012nd}
  J.~K.~Ahn {\it et al.} [RENO Collaboration],
  %``Observation of Reactor Electron Antineutrino Disappearance in the RENO Experiment,''
  Phys.\ Rev.\ Lett.\  {\bf 108}, 191802 (2012)
  doi:10.1103/PhysRevLett.108.191802
  [arXiv:1204.0626 [hep-ex]].
  %%CITATION = doi:10.1103/PhysRevLett.108.191802;%%
  %1205 citations counted in INSPIRE as of 13 Jan 2016



%\cite{Forero:2014bxa}
\bibitem{Forero:2014bxa}
  D.~V.~Forero, M.~Tortola and J.~W.~F.~Valle,
  %``Neutrino oscillations refitted,''
  Phys.\ Rev.\ D {\bf 90}, no. 9, 093006 (2014)
  doi:10.1103/PhysRevD.90.093006
  [arXiv:1405.7540 [hep-ph]].
  %%CITATION = doi:10.1103/PhysRevD.90.093006;%%
  %225 citations counted in INSPIRE as of 13 Jan 2016



%\cite{Agashe:2014kda}
\bibitem{Agashe:2014kda}
  K.~A.~Olive {\it et al.} [Particle Data Group Collaboration],
  %``Review of Particle Physics,''
  Chin.\ Phys.\ C {\bf 38}, 090001 (2014).
  doi:10.1088/1674-1137/38/9/090001
  %%CITATION = doi:10.1088/1674-1137/38/9/090001;%%
  %2720 citations counted in INSPIRE as of 13 Jan 2016



%\cite{Fritzsch:1977za}
\bibitem{Fritzsch:1977za}
  H.~Fritzsch,
  %``Calculating the Cabibbo Angle,''
  Phys.\ Lett.\ B {\bf 70}, 436 (1977).
  doi:10.1016/0370-2693(77)90408-7
  %%CITATION = doi:10.1016/0370-2693(77)90408-7;%%
  %528 citations counted in INSPIRE as of 13 Jan 2016



%\cite{Fukuyama:1997ky}
\bibitem{Fukuyama:1997ky}
  T.~Fukuyama and H.~Nishiura,
  %``Mass matrix of Majorana neutrinos,''
  hep-ph/9702253.
  %%CITATION = HEP-PH/9702253;%%
  %204 citations counted in INSPIRE as of 13 Jan 2016



%\cite{Du:1992iy}
\bibitem{Du:1992iy}
  D.~s.~Du and Z.~z.~Xing,
  %``A Modified Fritzsch ansatz with additional first order perturbation,''
  Phys.\ Rev.\ D {\bf 48}, 2349 (1993).
  doi:10.1103/PhysRevD.48.2349
  %%CITATION = doi:10.1103/PhysRevD.48.2349;%%
  %79 citations counted in INSPIRE as of 13 Jan 2016



%\cite{Barbieri:1994kw}
\bibitem{Barbieri:1994kw}
  R.~Barbieri, G.~R.~Dvali, A.~Strumia, Z.~Berezhiani and L.~J.~Hall,
  %``Flavor in supersymmetric grand unification: A Democratic approach,''
  Nucl.\ Phys.\ B {\bf 432}, 49 (1994)
  doi:10.1016/0550-3213(94)90593-2
  [hep-ph/9405428].
  %%CITATION = doi:10.1016/0550-3213(94)90593-2;%%
  %77 citations counted in INSPIRE as of 13 Jan 2016



%\cite{Peccei:1995fg}
\bibitem{Peccei:1995fg}
  R.~D.~Peccei and K.~Wang,
  %``Natural mass matrices,''
  Phys.\ Rev.\ D {\bf 53}, 2712 (1996)
  doi:10.1103/PhysRevD.53.2712
  [hep-ph/9509242].
  %%CITATION = doi:10.1103/PhysRevD.53.2712;%%
  %41 citations counted in INSPIRE as of 13 Jan 2016



%\cite{Fritzsch:1999ee}
\bibitem{Fritzsch:1999ee}
  H.~Fritzsch and Z.~z.~Xing,
  %``Mass and flavor mixing schemes of quarks and leptons,''
  Prog.\ Part.\ Nucl.\ Phys.\  {\bf 45}, 1 (2000)
  doi:10.1016/S0146-6410(00)00102-2
  [hep-ph/9912358].
  %%CITATION = doi:10.1016/S0146-6410(00)00102-2;%%
  %437 citations counted in INSPIRE as of 13 Jan 2016



%\cite{Roberts:2001zy}
\bibitem{Roberts:2001zy}
  R.~G.~Roberts, A.~Romanino, G.~G.~Ross and L.~Velasco-Sevilla,
  %``Precision test of a fermion mass texture,''
  Nucl.\ Phys.\ B {\bf 615}, 358 (2001)
  doi:10.1016/S0550-3213(01)00408-4
  [hep-ph/0104088].
  %%CITATION = doi:10.1016/S0550-3213(01)00408-4;%%
  %125 citations counted in INSPIRE as of 13 Jan 2016



%\cite{Nishiura:2002ei}
\bibitem{Nishiura:2002ei}
  H.~Nishiura, K.~Matsuda, T.~Kikuchi and T.~Fukuyama,
  %``Phenomenological analysis of lepton and quark mass matrices,''
  Phys.\ Rev.\ D {\bf 65}, 097301 (2002)
  doi:10.1103/PhysRevD.65.097301
  [hep-ph/0202189].
  %%CITATION = doi:10.1103/PhysRevD.65.097301;%%
  %21 citations counted in INSPIRE as of 13 Jan 2016



%\cite{deMedeirosVarzielas:2005ax}
\bibitem{deMedeirosVarzielas:2005ax}
  I.~de Medeiros Varzielas and G.~G.~Ross,
  %``SU(3) family symmetry and neutrino bi-tri-maximal mixing,''
  Nucl.\ Phys.\ B {\bf 733}, 31 (2006)
  doi:10.1016/j.nuclphysb.2005.10.039
  [hep-ph/0507176].
  %%CITATION = doi:10.1016/j.nuclphysb.2005.10.039;%%
  %193 citations counted in INSPIRE as of 13 Jan 2016



%\cite{Carcamo:2006dp}
\bibitem{Carcamo:2006dp}
  A.~E.~Carcamo Hernandez, R.~Martinez and J.~A.~Rodriguez,
  %``Different kind of textures of Yukawa coupling matrices in the two Higgs doublet model type III,''
  Eur.\ Phys.\ J.\ C {\bf 50}, 935 (2007)
  doi:10.1140/epjc/s10052-007-0264-0
  [hep-ph/0606190].
  %%CITATION = doi:10.1140/epjc/s10052-007-0264-0;%%
  %33 citations counted in INSPIRE as of 13 Jan 2016



%\cite{Kajiyama:2007gx}
\bibitem{Kajiyama:2007gx}
  Y.~Kajiyama, M.~Raidal and A.~Strumia,
  %``The Golden ratio prediction for the solar neutrino mixing,''
  Phys.\ Rev.\ D {\bf 76}, 117301 (2007)
  doi:10.1103/PhysRevD.76.117301
  [arXiv:0705.4559 [hep-ph]].
  %%CITATION = doi:10.1103/PhysRevD.76.117301;%%
  %107 citations counted in INSPIRE as of 13 Jan 2016



%\cite{CarcamoHernandez:2010im}
\bibitem{CarcamoHernandez:2010im}
  A.~E.~Carcamo Hernandez and R.~Rahman,
  %``Pseudo-Goldstone Higgs Doublets from Non-Vectorlike Grand Unified Higgs Sector,''
  arXiv:1007.0447 [hep-ph].
  %%CITATION = ARXIV:1007.0447;%%
  %13 citations counted in INSPIRE as of 13 Jan 2016



%\cite{Branco:2010tx}
\bibitem{Branco:2010tx}
  G.~C.~Branco, D.~Emmanuel-Costa and C.~Simoes,
  %``Nearest-Neighbour Interaction from an Abelian Symmetry and Deviations from Hermiticity,''
  Phys.\ Lett.\ B {\bf 690}, 62 (2010)
  doi:10.1016/j.physletb.2010.05.009
  [arXiv:1001.5065 [hep-ph]].
  %%CITATION = doi:10.1016/j.physletb.2010.05.009;%%
  %26 citations counted in INSPIRE as of 13 Jan 2016



%\cite{Leser:2011fz}
\bibitem{Leser:2011fz}
  P.~Leser and H.~Pas,
  %``Neutrino mass hierarchy and the origin of leptonic flavor mixing from the righthanded sector,''
  Phys.\ Rev.\ D {\bf 84}, 017303 (2011)
  doi:10.1103/PhysRevD.84.017303
  [arXiv:1104.2448 [hep-ph]].
  %%CITATION = doi:10.1103/PhysRevD.84.017303;%%
  %3 citations counted in INSPIRE as of 13 Jan 2016



%\cite{Gupta:2012dma}
\bibitem{Gupta:2012dma}
  M.~Gupta and G.~Ahuja,
  %``Flavor mixings and textures of the fermion mass matrices,''
  Int.\ J.\ Mod.\ Phys.\ A {\bf 27}, 1230033 (2012)
  doi:10.1142/S0217751X12300335
  [arXiv:1302.4823 [hep-ph]].
  %%CITATION = doi:10.1142/S0217751X12300335;%%
  %28 citations counted in INSPIRE as of 13 Jan 2016












%\cite{Hernandez:2013mcf}
\bibitem{Hernandez:2013mcf}
  A.~E.~Carcamo Hernandez, R.~Martinez and F.~Ochoa,
  %``Radiative seesaw-type mechanism of quark masses in $SU(3)_C \otimes SU(3)_L \otimes U(1)_X$,''
  Phys.\ Rev.\ D {\bf 87}, no. 7, 075009 (2013)
  doi:10.1103/PhysRevD.87.075009
  [arXiv:1302.1757 [hep-ph]].
  %%CITATION = doi:10.1103/PhysRevD.87.075009;%%
  %16 citations counted in INSPIRE as of 13 Jan 2016



%\cite{Pas:2014bra}
\bibitem{Pas:2014bra}
  H.~Päs and E.~Schumacher,
  %``Leptogenesis and $CP$ violation in $SU(5)$ models with lepton flavor mixing originating from the right-handed sector,''
  Phys.\ Rev.\ D {\bf 89}, no. 9, 096010 (2014)
  doi:10.1103/PhysRevD.89.096010
  [arXiv:1401.2328 [hep-ph]].
  %%CITATION = doi:10.1103/PhysRevD.89.096010;%%
  %2 citations counted in INSPIRE as of 13 Jan 2016



%\cite{Hernandez:2014hka}
\bibitem{Hernandez:2014hka}
  A.~E.~Carcamo Hernandez, S.~Kovalenko and I.~Schmidt,
  %``An SU(5) grand unified model with discrete flavour symmetries,''
  arXiv:1411.2913 [hep-ph].
  %%CITATION = ARXIV:1411.2913;%%
  %5 citations counted in INSPIRE as of 13 Jan 2016



%\cite{Hernandez:2014zsa}
\bibitem{Hernandez:2014zsa}
  A.~E.~C.~Hernández and I.~d.~M.~Varzielas,
  %``Viable textures for the fermion sector,''
  J.\ Phys.\ G {\bf 42}, no. 6, 065002 (2015)
  doi:10.1088/0954-3899/42/6/065002
  [arXiv:1410.2481 [hep-ph]].
  %%CITATION = doi:10.1088/0954-3899/42/6/065002;%%
  %8 citations counted in INSPIRE as of 13 Jan 2016



%\cite{Nishiura:2014psa}
\bibitem{Nishiura:2014psa}
  H.~Nishiura and T.~Fukuyama,
  %``Simple neutrino mass matrix with only two free parameters,''
  Mod.\ Phys.\ Lett.\ A {\bf 29}, 0147 (2014)
  doi:10.1142/S0217732314501478
  [arXiv:1405.2416 [hep-ph]].
  %%CITATION = doi:10.1142/S0217732314501478;%%
  %3 citations counted in INSPIRE as of 13 Jan 2016



%\cite{Frank:2014aca}
\bibitem{Frank:2014aca}
  M.~Frank, C.~Hamzaoui, N.~Pourtolami and M.~Toharia,
  %``Unified Flavor Symmetry from warped dimensions,''
  Phys.\ Lett.\ B {\bf 742}, 178 (2015)
  doi:10.1016/j.physletb.2015.01.025
  [arXiv:1406.2331 [hep-ph]].
  %%CITATION = doi:10.1016/j.physletb.2015.01.025;%%
  %2 citations counted in INSPIRE as of 13 Jan 2016



%\cite{Sinha:2015ooa}
\bibitem{Sinha:2015ooa}
  R.~Sinha, R.~Samanta and A.~Ghosal,
  %``Revisiting Allowed Two Zero Textures of Neutrino Mass Matrices through Linear and Inverse Seesaw,''
  arXiv:1508.05227 [hep-ph].
  %%CITATION = ARXIV:1508.05227;%%
  %2 citations counted in INSPIRE as of 13 Jan 2016



%\cite{Nishiura:2015qia}
\bibitem{Nishiura:2015qia}
  H.~Nishiura and T.~Fukuyama,
  %``Simple mass matrices of neutrinos and quarks consistent with observed mixings and masses,''
  Phys.\ Lett.\ B {\bf 753}, 57 (2016)
  doi:10.1016/j.physletb.2015.11.080
  [arXiv:1510.01035 [hep-ph]].
  %%CITATION = doi:10.1016/j.physletb.2015.11.080;%%
  %1 citations counted in INSPIRE as of 13 Jan 2016



%\cite{Gautam:2015kya}
\bibitem{Gautam:2015kya}
  R.~R.~Gautam, M.~Singh and M.~Gupta,
  %``Neutrino mass matrices with one texture zero and a vanishing neutrino mass,''
  Phys.\ Rev.\ D {\bf 92}, no. 1, 013006 (2015)
  doi:10.1103/PhysRevD.92.013006
  [arXiv:1506.04868 [hep-ph]].
  %%CITATION = doi:10.1103/PhysRevD.92.013006;%%
  %5 citations counted in INSPIRE as of 13 Jan 2016



%\cite{Pas:2015hca}
\bibitem{Pas:2015hca}
  H.~Päs and E.~Schumacher,
  %``Common origin of $R_K$ and neutrino masses,''
  Phys.\ Rev.\ D {\bf 92}, no. 11, 114025 (2015)
  doi:10.1103/PhysRevD.92.114025
  [arXiv:1510.08757 [hep-ph]].
  %%CITATION = doi:10.1103/PhysRevD.92.114025;%%
  %2 citations counted in INSPIRE as of 13 Jan 2016





%\cite{Ishimori:2010au}
\bibitem{Ishimori:2010au}
  H.~Ishimori, T.~Kobayashi, H.~Ohki, Y.~Shimizu, H.~Okada and M.~Tanimoto,
  %``Non-Abelian Discrete Symmetries in Particle Physics,''
  Prog.\ Theor.\ Phys.\ Suppl.\  {\bf 183}, 1 (2010)
  doi:10.1143/PTPS.183.1
  [arXiv:1003.3552 [hep-th]].
  %%CITATION = doi:10.1143/PTPS.183.1;%%
  %379 citations counted in INSPIRE as of 13 Jan 2016



%\cite{Altarelli:2010gt}
\bibitem{Altarelli:2010gt}
  G.~Altarelli and F.~Feruglio,
  %``Discrete Flavor Symmetries and Models of Neutrino Mixing,''
  Rev.\ Mod.\ Phys.\  {\bf 82}, 2701 (2010)
  doi:10.1103/RevModPhys.82.2701
  [arXiv:1002.0211 [hep-ph]].
  %%CITATION = doi:10.1103/RevModPhys.82.2701;%%
  %429 citations counted in INSPIRE as of 13 Jan 2016







%\cite{King:2013eh}
\bibitem{King:2013eh}
  S.~F.~King and C.~Luhn,
  %``Neutrino Mass and Mixing with Discrete Symmetry,''
  Rept.\ Prog.\ Phys.\  {\bf 76}, 056201 (2013)
  doi:10.1088/0034-4885/76/5/056201
  [arXiv:1301.1340 [hep-ph]].
  %%CITATION = doi:10.1088/0034-4885/76/5/056201;%%
  %248 citations counted in INSPIRE as of 13 Jan 2016






%\cite{King:2014nza}
\bibitem{King:2014nza}
  S.~F.~King, A.~Merle, S.~Morisi, Y.~Shimizu and M.~Tanimoto,
  %``Neutrino Mass and Mixing: from Theory to Experiment,''
  New J.\ Phys.\  {\bf 16}, 045018 (2014)
  doi:10.1088/1367-2630/16/4/045018
  [arXiv:1402.4271 [hep-ph]].
  %%CITATION = doi:10.1088/1367-2630/16/4/045018;%%
  %91 citations counted in INSPIRE as of 13 Jan 2016










%\cite{Ma:2001dn}
\bibitem{Ma:2001dn}
  E.~Ma and G.~Rajasekaran,
  %``Softly broken A(4) symmetry for nearly degenerate neutrino masses,''
  Phys.\ Rev.\ D {\bf 64}, 113012 (2001)
  doi:10.1103/PhysRevD.64.113012
  [hep-ph/0106291].
  %%CITATION = doi:10.1103/PhysRevD.64.113012;%%
  %560 citations counted in INSPIRE as of 13 Jan 2016



%\cite{He:2006dk}
\bibitem{He:2006dk}
  X.~G.~He, Y.~Y.~Keum and R.~R.~Volkas,
  %``A(4) flavor symmetry breaking scheme for understanding quark and neutrino mixing angles,''
  JHEP {\bf 0604}, 039 (2006)
  doi:10.1088/1126-6708/2006/04/039
  [hep-ph/0601001].
  %%CITATION = doi:10.1088/1126-6708/2006/04/039;%%
  %225 citations counted in INSPIRE as of 13 Jan 2016



%\cite{Chen:2009um}
\bibitem{Chen:2009um}
  M.~C.~Chen and S.~F.~King,
  %``A4 See-Saw Models and Form Dominance,''
  JHEP {\bf 0906}, 072 (2009)
  doi:10.1088/1126-6708/2009/06/072
  [arXiv:0903.0125 [hep-ph]].
  %%CITATION = doi:10.1088/1126-6708/2009/06/072;%%
  %106 citations counted in INSPIRE as of 13 Jan 2016



%\cite{Dong:2010gk}
\bibitem{Dong:2010gk}
  P.~V.~Dong, L.~T.~Hue, H.~N.~Long and D.~V.~Soa,
  %``The 3-3-1 model with A_4 flavor symmetry,''
  Phys.\ Rev.\ D {\bf 81}, 053004 (2010)
  doi:10.1103/PhysRevD.81.053004
  [arXiv:1001.4625 [hep-ph]].
  %%CITATION = doi:10.1103/PhysRevD.81.053004;%%
  %34 citations counted in INSPIRE as of 13 Jan 2016



%\cite{Ahn:2012tv}
\bibitem{Ahn:2012tv}
  Y.~H.~Ahn and S.~K.~Kang,
  %``Non-zero $\theta_{13}$ and CP violation in a model with $A_4$ flavor symmetry,''
  Phys.\ Rev.\ D {\bf 86}, 093003 (2012)
  doi:10.1103/PhysRevD.86.093003
  [arXiv:1203.4185 [hep-ph]].
  %%CITATION = doi:10.1103/PhysRevD.86.093003;%%
  %36 citations counted in INSPIRE as of 13 Jan 2016



%\cite{Memenga:2013vc}
\bibitem{Memenga:2013vc}
  N.~Memenga, W.~Rodejohann and H.~Zhang,
  %``$A_4$ flavor symmetry model for Dirac neutrinos and sizable $U_{e3}$,''
  Phys.\ Rev.\ D {\bf 87}, no. 5, 053021 (2013)
  doi:10.1103/PhysRevD.87.053021
  [arXiv:1301.2963 [hep-ph]].
  %%CITATION = doi:10.1103/PhysRevD.87.053021;%%
  %18 citations counted in INSPIRE as of 13 Jan 2016



%\cite{Felipe:2013vwa}
\bibitem{Felipe:2013vwa}
  R.~Gonzalez Felipe, H.~Serodio and J.~P.~Silva,
  %``Neutrino masses and mixing in A4 models with three Higgs doublets,''
  Phys.\ Rev.\ D {\bf 88}, no. 1, 015015 (2013)
  doi:10.1103/PhysRevD.88.015015
  [arXiv:1304.3468 [hep-ph]].
  %%CITATION = doi:10.1103/PhysRevD.88.015015;%%
  %13 citations counted in INSPIRE as of 13 Jan 2016



%\cite{Varzielas:2012ai}
\bibitem{Varzielas:2012ai}
  I.~de Medeiros Varzielas and D.~Pidt,
  %``UV completions of flavour models and large theta_{13},''
  JHEP {\bf 1303}, 065 (2013)
  doi:10.1007/JHEP03(2013)065
  [arXiv:1211.5370 [hep-ph]].
  %%CITATION = doi:10.1007/JHEP03(2013)065;%%
  %16 citations counted in INSPIRE as of 13 Jan 2016



%\cite{Ishimori:2012fg}
\bibitem{Ishimori:2012fg}
  H.~Ishimori and E.~Ma,
  %``New Simple $A_4$ Neutrino Model for Nonzero $\theta_{13}$ and Large $\delta_{CP}$,''
  Phys.\ Rev.\ D {\bf 86}, 045030 (2012)
  doi:10.1103/PhysRevD.86.045030
  [arXiv:1205.0075 [hep-ph]].
  %%CITATION = doi:10.1103/PhysRevD.86.045030;%%
  %41 citations counted in INSPIRE as of 13 Jan 2016



%\cite{King:2013hj}
\bibitem{King:2013hj}
  S.~F.~King, S.~Morisi, E.~Peinado and J.~W.~F.~Valle,
  %``Quark-Lepton Mass Relation in a Realistic $A_4$ Extension of the Standard Model,''
  Phys.\ Lett.\ B {\bf 724}, 68 (2013)
  doi:10.1016/j.physletb.2013.05.067
  [arXiv:1301.7065 [hep-ph]].
  %%CITATION = doi:10.1016/j.physletb.2013.05.067;%%
  %19 citations counted in INSPIRE as of 13 Jan 2016



%\cite{Hernandez:2013dta}
\bibitem{Hernandez:2013dta}
  A.~E.~Carcamo Hernandez, I.~de Medeiros Varzielas, S.~G.~Kovalenko, H.~Päs and I.~Schmidt,
  %``Lepton masses and mixings in an $A_4$ multi-Higgs model with a radiative seesaw mechanism,''
  Phys.\ Rev.\ D {\bf 88}, no. 7, 076014 (2013)
  doi:10.1103/PhysRevD.88.076014
  [arXiv:1307.6499 [hep-ph]].
  %%CITATION = doi:10.1103/PhysRevD.88.076014;%%
  %31 citations counted in INSPIRE as of 13 Jan 2016



%\cite{Babu:2002dz}
\bibitem{Babu:2002dz}
  K.~S.~Babu, E.~Ma and J.~W.~F.~Valle,
  %``Underlying A(4) symmetry for the neutrino mass matrix and the quark mixing matrix,''
  Phys.\ Lett.\ B {\bf 552}, 207 (2003)
  doi:10.1016/S0370-2693(02)03153-2
  [hep-ph/0206292].
  %%CITATION = doi:10.1016/S0370-2693(02)03153-2;%%
  %556 citations counted in INSPIRE as of 13 Jan 2016



%\cite{Altarelli:2005yx}
\bibitem{Altarelli:2005yx}
  G.~Altarelli and F.~Feruglio,
  %``Tri-bimaximal neutrino mixing, A(4) and the modular symmetry,''
  Nucl.\ Phys.\ B {\bf 741}, 215 (2006)
  doi:10.1016/j.nuclphysb.2006.02.015
  [hep-ph/0512103].
  %%CITATION = doi:10.1016/j.nuclphysb.2006.02.015;%%
  %452 citations counted in INSPIRE as of 13 Jan 2016



%\cite{Morisi:2013eca}
\bibitem{Morisi:2013eca}
  S.~Morisi, M.~Nebot, K.~M.~Patel, E.~Peinado and J.~W.~F.~Valle,
  %``Quark-Lepton Mass Relation and CKM mixing in an A4 Extension of the Minimal Supersymmetric Standard Model,''
  Phys.\ Rev.\ D {\bf 88}, 036001 (2013)
  doi:10.1103/PhysRevD.88.036001
  [arXiv:1303.4394 [hep-ph]].
  %%CITATION = doi:10.1103/PhysRevD.88.036001;%%
  %12 citations counted in INSPIRE as of 13 Jan 2016



%\cite{Altarelli:2005yp}
\bibitem{Altarelli:2005yp}
  G.~Altarelli and F.~Feruglio,
  %``Tri-bimaximal neutrino mixing from discrete symmetry in extra dimensions,''
  Nucl.\ Phys.\ B {\bf 720}, 64 (2005)
  doi:10.1016/j.nuclphysb.2005.05.005
  [hep-ph/0504165].
  %%CITATION = doi:10.1016/j.nuclphysb.2005.05.005;%%
  %461 citations counted in INSPIRE as of 13 Jan 2016



%\cite{Kadosh:2010rm}
\bibitem{Kadosh:2010rm}
  A.~Kadosh and E.~Pallante,
  %``An A(4) flavor model for quarks and leptons in warped geometry,''
  JHEP {\bf 1008}, 115 (2010)
  doi:10.1007/JHEP08(2010)115
  [arXiv:1004.0321 [hep-ph]].
  %%CITATION = doi:10.1007/JHEP08(2010)115;%%
  %42 citations counted in INSPIRE as of 13 Jan 2016



%\cite{Kadosh:2013nra}
\bibitem{Kadosh:2013nra}
  A.~Kadosh,
  %``$\Theta_13$ and charged Lepton Flavor Violation in "warped" $A_4$ models,''
  JHEP {\bf 1306}, 114 (2013)
  doi:10.1007/JHEP06(2013)114
  [arXiv:1303.2645 [hep-ph]].
  %%CITATION = doi:10.1007/JHEP06(2013)114;%%
  %5 citations counted in INSPIRE as of 13 Jan 2016



%\cite{delAguila:2010vg}
\bibitem{delAguila:2010vg}
  F.~del Aguila, A.~Carmona and J.~Santiago,
  %``Neutrino Masses from an A4 Symmetry in Holographic Composite Higgs Models,''
  JHEP {\bf 1008}, 127 (2010)
  doi:10.1007/JHEP08(2010)127
  [arXiv:1001.5151 [hep-ph]].
  %%CITATION = doi:10.1007/JHEP08(2010)127;%%
  %62 citations counted in INSPIRE as of 13 Jan 2016



%\cite{Campos:2014lla}
\bibitem{Campos:2014lla}
  M.~D.~Campos, A.~E.~Cárcamo Hernández, S.~Kovalenko, I.~Schmidt and E.~Schumacher,
  %``Fermion masses and mixings in an $SU(5)$ grand unified model with an extra flavor symmetry,''
  Phys.\ Rev.\ D {\bf 90}, no. 1, 016006 (2014)
  doi:10.1103/PhysRevD.90.016006
  [arXiv:1403.2525 [hep-ph]].
  %%CITATION = doi:10.1103/PhysRevD.90.016006;%%
  %14 citations counted in INSPIRE as of 13 Jan 2016



%\cite{Vien:2014pta}
\bibitem{Vien:2014pta}
  V.~V.~Vien and H.~N.~Long,
  %``Neutrino mixing with nonzero $\theta_{13}$ and CP violation in the 3-3-1 model based on $A_4$ flavor symmetry,''
  Int.\ J.\ Mod.\ Phys.\ A {\bf 30}, no. 21, 1550117 (2015)
  doi:10.1142/S0217751X15501171
  [arXiv:1405.4665 [hep-ph]].
  %%CITATION = doi:10.1142/S0217751X15501171;%%
  %4 citations counted in INSPIRE as of 13 Jan 2016



%\cite{Hernandez:2015tna}
\bibitem{Hernandez:2015tna}
  A.~E.~Cárcamo Hernández and R.~Martinez,
  %``A predictive $331$ model with $A_{4}$ flavour symmetry,''
  Nucl. Phys. B 905 (2016) 337
  doi:10.1016/j.nuclphysb.2016.02.025
  arXiv:1501.05937 [hep-ph].
  %%CITATION = ARXIV:1501.05937;%%
  %10 citations counted in INSPIRE as of 13 Jan 2016



%\cite{Nishi:2016jqg}
\bibitem{Nishi:2016jqg}
  C.~C.~Nishi, \textbf{	Phys. Rev. D 93, 093009 (2016), }
  %``A new and trivial CP symmetry for extended $A_4$ flavor,''
  arXiv:1601.00977 [hep-ph].
  %%CITATION = ARXIV:1601.00977;%%




%\cite{Hernandez:2015hrt}
\bibitem{Hernandez:2015hrt}
  A.~E.~C.~Hernández,  \textbf{Eur.Phys.J. C76 (2016) no.9, 503,}
  %``The 750 GeV diphoton resonance can cause the SM fermion mass and mixing pattern,''
  arXiv:1512.09092 [hep-ph].
  %%CITATION = ARXIV:1512.09092;%%
  %14 citations counted in INSPIRE as of 13 Jan 2016




%\cite{Chen:2004rr}
\bibitem{Chen:2004rr}
  S.~L.~Chen, M.~Frigerio and E.~Ma,
  %``Large neutrino mixing and normal mass hierarchy: A Discrete understanding,''
  Phys.\ Rev.\ D {\bf 70}, 073008 (2004)
  [Phys.\ Rev.\ D {\bf 70}, 079905 (2004)]
  doi:10.1103/PhysRevD.70.079905, 10.1103/PhysRevD.70.073008
  [hep-ph/0404084].
  %%CITATION = doi:10.1103/PhysRevD.70.079905, 10.1103/PhysRevD.70.073008;%%
  %122 citations counted in INSPIRE as of 13 Jan 2016



%\cite{Dong:2011vb}
\bibitem{Dong:2011vb}
  P.~V.~Dong, H.~N.~Long, C.~H.~Nam and V.~V.~Vien,
  %``The $S_3$ flavor symmetry in 3-3-1 models,''
  Phys.\ Rev.\ D {\bf 85}, 053001 (2012)
  doi:10.1103/PhysRevD.85.053001
  [arXiv:1111.6360 [hep-ph]].
  %%CITATION = doi:10.1103/PhysRevD.85.053001;%%
  %39 citations counted in INSPIRE as of 13 Jan 2016



%\cite{Bhattacharyya:2010hp}
\bibitem{Bhattacharyya:2010hp}
  G.~Bhattacharyya, P.~Leser and H.~Pas,
  %``Exotic Higgs boson decay modes as a harbinger of $S_3$ flavor symmetry,''
  Phys.\ Rev.\ D {\bf 83}, 011701 (2011)
  doi:10.1103/PhysRevD.83.011701
  [arXiv:1006.5597 [hep-ph]].
  %%CITATION = doi:10.1103/PhysRevD.83.011701;%%
  %43 citations counted in INSPIRE as of 13 Jan 2016



%\cite{Dias:2012bh}
\bibitem{Dias:2012bh}
  A.~G.~Dias, A.~C.~B.~Machado and C.~C.~Nishi,
  %``An $S_3$ Model for Lepton Mass Matrices with Nearly Minimal Texture,''
  Phys.\ Rev.\ D {\bf 86}, 093005 (2012)
  doi:10.1103/PhysRevD.86.093005
  [arXiv:1206.6362 [hep-ph]].
  %%CITATION = doi:10.1103/PhysRevD.86.093005;%%
  %11 citations counted in INSPIRE as of 13 Jan 2016



%\cite{Meloni:2012ci}
\bibitem{Meloni:2012ci}
  D.~Meloni,
  %``$S_3$ as a flavour symmetry for quarks and leptons after the Daya Bay result on $\theta_{13}$,''
  JHEP {\bf 1205}, 124 (2012)
  doi:10.1007/JHEP05(2012)124
  [arXiv:1203.3126 [hep-ph]].
  %%CITATION = doi:10.1007/JHEP05(2012)124;%%
  %19 citations counted in INSPIRE as of 13 Jan 2016



%\cite{Canales:2013cga}
\bibitem{Canales:2013cga}
  F.~González Canales, A.~Mondragón, M.~Mondragón, U.~J.~Saldaña Salazar and L.~Velasco-Sevilla,
  %``Quark sector of S3 models: classification and comparison with experimental data,''
  Phys.\ Rev.\ D {\bf 88}, 096004 (2013)
  doi:10.1103/PhysRevD.88.096004
  [arXiv:1304.6644 [hep-ph]].
  %%CITATION = doi:10.1103/PhysRevD.88.096004;%%
  %25 citations counted in INSPIRE as of 13 Jan 2016



%\cite{Ma:2013zca}
\bibitem{Ma:2013zca}
  E.~Ma and B.~Melic,
  %``Updated $S_{3}$ model of quarks,''
  Phys.\ Lett.\ B {\bf 725}, 402 (2013)
  doi:10.1016/j.physletb.2013.07.015
  [arXiv:1303.6928 [hep-ph]].
  %%CITATION = doi:10.1016/j.physletb.2013.07.015;%%
  %19 citations counted in INSPIRE as of 13 Jan 2016



%\cite{Kajiyama:2013sza}
\bibitem{Kajiyama:2013sza}
  Y.~Kajiyama, H.~Okada and K.~Yagyu,
  %``Electron/Muon Specific Two Higgs Doublet Model,''
  Nucl.\ Phys.\ B {\bf 887}, 358 (2014)
  doi:10.1016/j.nuclphysb.2014.08.009
  [arXiv:1309.6234 [hep-ph]].
  %%CITATION = doi:10.1016/j.nuclphysb.2014.08.009;%%
  %28 citations counted in INSPIRE as of 13 Jan 2016



%\cite{Hernandez:2013hea}
\bibitem{Hernandez:2013hea}
  A.~E.~Cárcamo Hernández, R.~Martínez and F.~Ochoa,
  %``Fermion masses and mixings in a minimal 331 model with S3 flavor symmetry,''
  arXiv:1309.6567 [hep-ph].
  %%CITATION = ARXIV:1309.6567;%%
  %26 citations counted in INSPIRE as of 13 Jan 2016



%\cite{Ma:2014qra}
\bibitem{Ma:2014qra}
  E.~Ma and R.~Srivastava,
  %``Dirac or inverse seesaw neutrino masses with $B-L$ gauge symmetry and $S_3$ flavor symmetry,''
  Phys.\ Lett.\ B {\bf 741}, 217 (2015)
  doi:10.1016/j.physletb.2014.12.049
  [arXiv:1411.5042 [hep-ph]].
  %%CITATION = doi:10.1016/j.physletb.2014.12.049;%%
  %12 citations counted in INSPIRE as of 13 Jan 2016



%\cite{Hernandez:2014vta}
\bibitem{Hernandez:2014vta}
  A.~E.~C.~Hernández, R.~Martinez and J.~Nisperuza,
  %``$S_3$ discrete group as a source of the quark mass and mixing pattern in $331$ models,''
  Eur.\ Phys.\ J.\ C {\bf 75}, no. 2, 72 (2015)
  doi:10.1140/epjc/s10052-015-3278-z
  [arXiv:1401.0937 [hep-ph]].
  %%CITATION = doi:10.1140/epjc/s10052-015-3278-z;%%
  %20 citations counted in INSPIRE as of 13 Jan 2016



%\cite{Hernandez:2014lpa}
\bibitem{Hernandez:2014lpa}
  A.~E.~C.~Hernández, E.~C.~Mur and R.~Martinez,
  %``Lepton masses and mixing in $SU(3)_{C}\otimes SU(3)_{L}\otimes U(1)_{X}$ models with a $S_3$ flavor symmetry,''
  Phys.\ Rev.\ D {\bf 90}, no. 7, 073001 (2014)
  doi:10.1103/PhysRevD.90.073001
  [arXiv:1407.5217 [hep-ph]].
  %%CITATION = doi:10.1103/PhysRevD.90.073001;%%
  %13 citations counted in INSPIRE as of 13 Jan 2016



%\cite{Hernandez:2015dga}
\bibitem{Hernandez:2015dga}
  A.~E.~C.~Hernández, I.~d.~M.~Varzielas and E.~Schumacher,
  %``Fermion and scalar phenomenology of a 2-Higgs doublet model with $S_3$,''
  Phys. \ Rev. \ D\  93, 016003 (2016)
doi:	10.1103/PhysRevD.93.016003
  arXiv:1509.02083 [hep-ph].
  %%CITATION = ARXIV:1509.02083;%%
  %6 citations counted in INSPIRE as of 13 Jan 2016



%\cite{Hernandez:2015zeh}
\bibitem{Hernandez:2015zeh}
  A.~E.~C.~Hernández, I.~d.~M.~Varzielas and N.~A.~Neill,
  %``A Novel Randall-Sundrum Model with $S_{3}$ Flavor Symmetry,''
  arXiv:1511.07420 [hep-ph].
  %%CITATION = ARXIV:1511.07420;%%
  %2 citations counted in INSPIRE as of 13 Jan 2016



%\cite{Hernandez:2016rbi}
\bibitem{Hernandez:2016rbi}
  A.~E.~C.~Hernández, I.~d.~M.~Varzielas and E.~Schumacher,
  %``The $750\,\text{GeV}$ diphoton resonance in the light of a 2HDM with $S_3$ flavour symmetry,''
  arXiv:1601.00661 [hep-ph].
  %%CITATION = ARXIV:1601.00661;%%
  %6 citations counted in INSPIRE as of 13 Jan 2016



%\cite{Mohapatra:2012tb}
\bibitem{Mohapatra:2012tb}
  R.~N.~Mohapatra and C.~C.~Nishi,
  %``$S_4$ Flavored CP Symmetry for Neutrinos,''
  Phys.\ Rev.\ D {\bf 86}, 073007 (2012)
  doi:10.1103/PhysRevD.86.073007
  [arXiv:1208.2875 [hep-ph]].
  %%CITATION = doi:10.1103/PhysRevD.86.073007;%%
  %56 citations counted in INSPIRE as of 13 Jan 2016



%\cite{Varzielas:2012pa}
\bibitem{Varzielas:2012pa}
  I.~de Medeiros Varzielas and L.~Lavoura,
  %``Flavour models for $TM_{1}$ lepton mixing,''
  J.\ Phys.\ G {\bf 40}, 085002 (2013)
  doi:10.1088/0954-3899/40/8/085002
  [arXiv:1212.3247 [hep-ph]].
  %%CITATION = doi:10.1088/0954-3899/40/8/085002;%%
  %26 citations counted in INSPIRE as of 13 Jan 2016



%\cite{Ding:2013hpa}
\bibitem{Ding:2013hpa}
  G.~J.~Ding, S.~F.~King, C.~Luhn and A.~J.~Stuart,
  %``Spontaneous CP violation from vacuum alignment in $S_4$ models of leptons,''
  JHEP {\bf 1305}, 084 (2013)
  doi:10.1007/JHEP05(2013)084
  [arXiv:1303.6180 [hep-ph]].
  %%CITATION = doi:10.1007/JHEP05(2013)084;%%
  %50 citations counted in INSPIRE as of 13 Jan 2016



%\cite{Ishimori:2010fs}
\bibitem{Ishimori:2010fs}
  H.~Ishimori, Y.~Shimizu, M.~Tanimoto and A.~Watanabe,
  %``Neutrino masses and mixing from $S_{4}$ flavor twisting,''
  Phys.\ Rev.\ D {\bf 83}, 033004 (2011)
  doi:10.1103/PhysRevD.83.033004
  [arXiv:1010.3805 [hep-ph]].
  %%CITATION = doi:10.1103/PhysRevD.83.033004;%%
  %41 citations counted in INSPIRE as of 13 Jan 2016



%\cite{Ding:2013eca}
\bibitem{Ding:2013eca}
  G.~J.~Ding and Y.~L.~Zhou,
  %``Dirac Neutrinos with $S_4$ Flavor Symmetry in Warped Extra Dimensions,''
  Nucl.\ Phys.\ B {\bf 876}, 418 (2013)
  doi:10.1016/j.nuclphysb.2013.08.011
  [arXiv:1304.2645 [hep-ph]].
  %%CITATION = doi:10.1016/j.nuclphysb.2013.08.011;%%
  %14 citations counted in INSPIRE as of 13 Jan 2016



%\cite{Hagedorn:2011un}
\bibitem{Hagedorn:2011un}
  C.~Hagedorn and M.~Serone,
  %``Leptons in Holographic Composite Higgs Models with Non-Abelian Discrete Symmetries,''
  JHEP {\bf 1110}, 083 (2011)
  doi:10.1007/JHEP10(2011)083
  [arXiv:1106.4021 [hep-ph]].
  %%CITATION = doi:10.1007/JHEP10(2011)083;%%
  %17 citations counted in INSPIRE as of 13 Jan 2016



%\cite{Campos:2014zaa}
\bibitem{Campos:2014zaa}
  M.~D.~Campos, A.~E.~C.~Hernández, H.~Päs and E.~Schumacher,
  %``Higgs $\rightarrow$ $\mu\tau$ as an indication for $S_4$ flavor symmetry,''
  Phys.\ Rev.\ D {\bf 91}, no. 11, 116011 (2015)
  doi:10.1103/PhysRevD.91.116011
  [arXiv:1408.1652 [hep-ph]].
  %%CITATION = doi:10.1103/PhysRevD.91.116011;%%
  %36 citations counted in INSPIRE as of 13 Jan 2016



%\cite{Dong:2010zu}
\bibitem{Dong:2010zu}
  P.~V.~Dong, H.~N.~Long, D.~V.~Soa and V.~V.~Vien,
  %``The 3-3-1 model with $S_4$ flavor symmetry,''
  Eur.\ Phys.\ J.\ C {\bf 71}, 1544 (2011)
  doi:10.1140/epjc/s10052-011-1544-2
  [arXiv:1009.2328 [hep-ph]].
  %%CITATION = doi:10.1140/epjc/s10052-011-1544-2;%%
  %38 citations counted in INSPIRE as of 13 Jan 2016



%\cite{VanVien:2015xha}
\bibitem{VanVien:2015xha}
  V.~V.~Vien, H.~N.~Long and D.~P.~Khoi,
  %``Neutrino Mixing with Non-Zero $\theta_{13}$ and CP Violation in the 3-3-1 Model Based on $S_4$ Flavor Symmetry,''
  Int.\ J.\ Mod.\ Phys.\ A {\bf 30}, no. 17, 1550102 (2015)
  doi:10.1142/S0217751X1550102X
  [arXiv:1506.06063 [hep-ph]].
  %%CITATION = doi:10.1142/S0217751X1550102X;%%



%\cite{Frampton:1994rk}
\bibitem{Frampton:1994rk}
  P.~H.~Frampton and T.~W.~Kephart,
  %``Simple nonAbelian finite flavor groups and fermion masses,''
  Int.\ J.\ Mod.\ Phys.\ A {\bf 10}, 4689 (1995)
  doi:10.1142/S0217751X95002187
  [hep-ph/9409330].
  %%CITATION = doi:10.1142/S0217751X95002187;%%
  %143 citations counted in INSPIRE as of 13 Jan 2016



%\cite{Grimus:2003kq}
\bibitem{Grimus:2003kq}
  W.~Grimus and L.~Lavoura,
  %``A Discrete symmetry group for maximal atmospheric neutrino mixing,''
  Phys.\ Lett.\ B {\bf 572}, 189 (2003)
  doi:10.1016/j.physletb.2003.08.032
  [hep-ph/0305046].
  %%CITATION = doi:10.1016/j.physletb.2003.08.032;%%
  %233 citations counted in INSPIRE as of 13 Jan 2016



%\cite{Grimus:2004rj}
\bibitem{Grimus:2004rj}
  W.~Grimus, A.~S.~Joshipura, S.~Kaneko, L.~Lavoura and M.~Tanimoto,
  %``Lepton mixing angle $\theta_{13} = 0$ with a horizontal symmetry $D_4$,''
  JHEP {\bf 0407}, 078 (2004)
  doi:10.1088/1126-6708/2004/07/078
  [hep-ph/0407112].
  %%CITATION = doi:10.1088/1126-6708/2004/07/078;%%
  %111 citations counted in INSPIRE as of 13 Jan 2016



%\cite{Frigerio:2004jg}
\bibitem{Frigerio:2004jg}
  M.~Frigerio, S.~Kaneko, E.~Ma and M.~Tanimoto,
  %``Quaternion family symmetry of quarks and leptons,''
  Phys.\ Rev.\ D {\bf 71}, 011901 (2005)
  doi:10.1103/PhysRevD.71.011901
  [hep-ph/0409187].
  %%CITATION = doi:10.1103/PhysRevD.71.011901;%%
  %75 citations counted in INSPIRE as of 13 Jan 2016



%\cite{Babu:2004tn}
\bibitem{Babu:2004tn}
  K.~S.~Babu and J.~Kubo,
  %``Dihedral families of quarks, leptons and Higgses,''
  Phys.\ Rev.\ D {\bf 71}, 056006 (2005)
  doi:10.1103/PhysRevD.71.056006
  [hep-ph/0411226].
  %%CITATION = doi:10.1103/PhysRevD.71.056006;%%
  %115 citations counted in INSPIRE as of 13 Jan 2016



%\cite{Adulpravitchai:2008yp}
\bibitem{Adulpravitchai:2008yp}
  A.~Adulpravitchai, A.~Blum and C.~Hagedorn,
  %``A Supersymmetric D4 Model for mu-tau Symmetry,''
  JHEP {\bf 0903}, 046 (2009)
  doi:10.1088/1126-6708/2009/03/046
  [arXiv:0812.3799 [hep-ph]].
  %%CITATION = doi:10.1088/1126-6708/2009/03/046;%%
  %40 citations counted in INSPIRE as of 13 Jan 2016



%\cite{Ishimori:2008gp}
\bibitem{Ishimori:2008gp}
  H.~Ishimori, T.~Kobayashi, H.~Ohki, Y.~Omura, R.~Takahashi and M.~Tanimoto,
  %``D(4) Flavor Symmetry for Neutrino Masses and Mixing,''
  Phys.\ Lett.\ B {\bf 662}, 178 (2008)
  doi:10.1016/j.physletb.2008.03.007
  [arXiv:0802.2310 [hep-ph]].
  %%CITATION = doi:10.1016/j.physletb.2008.03.007;%%
  %37 citations counted in INSPIRE as of 13 Jan 2016



%\cite{Hagedorn:2010mq}
\bibitem{Hagedorn:2010mq}
  C.~Hagedorn and R.~Ziegler,
  %``$\mu-\tau$ Symmetry and Charged Lepton Mass Hierarchy in a Supersymmetric $D_4$ Model,''
  Phys.\ Rev.\ D {\bf 82}, 053011 (2010)
  doi:10.1103/PhysRevD.82.053011
  [arXiv:1007.1888 [hep-ph]].
  %%CITATION = doi:10.1103/PhysRevD.82.053011;%%
  %9 citations counted in INSPIRE as of 13 Jan 2016



%\cite{Meloni:2011cc}
\bibitem{Meloni:2011cc}
  D.~Meloni, S.~Morisi and E.~Peinado,
  %``Stability of dark matter from the D4xZ2 flavor group,''
  Phys.\ Lett.\ B {\bf 703}, 281 (2011)
  doi:10.1016/j.physletb.2011.07.084
  [arXiv:1104.0178 [hep-ph]].
  %%CITATION = doi:10.1016/j.physletb.2011.07.084;%%
  %19 citations counted in INSPIRE as of 13 Jan 2016



%\cite{Vien:2013zra}
\bibitem{Vien:2013zra}
  V.~V.~Vien and H.~N.~Long,
  %``The $D_4$ flavor symmery in 3-3-1 model with neutral leptons,''
  Int.\ J.\ Mod.\ Phys.\ A {\bf 28}, 1350159 (2013)
  doi:10.1142/S0217751X13501595
  [arXiv:1312.5034 [hep-ph]].
  %%CITATION = doi:10.1142/S0217751X13501595;%%
  %14 citations counted in INSPIRE as of 13 Jan 2016



%\cite{Luhn:2007sy}
\bibitem{Luhn:2007sy}
  C.~Luhn, S.~Nasri and P.~Ramond,
  %``Tri-bimaximal neutrino mixing and the family symmetry semidirect product of Z(7) and Z(3),''
  Phys.\ Lett.\ B {\bf 652}, 27 (2007)
  doi:10.1016/j.physletb.2007.06.059
  [arXiv:0706.2341 [hep-ph]].
  %%CITATION = doi:10.1016/j.physletb.2007.06.059;%%
  %136 citations counted in INSPIRE as of 13 Jan 2016



%\cite{Hagedorn:2008bc}
\bibitem{Hagedorn:2008bc}
  C.~Hagedorn, M.~A.~Schmidt and A.~Y.~Smirnov,
  %``Lepton Mixing and Cancellation of the Dirac Mass Hierarchy in SO(10) GUTs with Flavor Symmetries T(7) and Sigma(81),''
  Phys.\ Rev.\ D {\bf 79}, 036002 (2009)
  doi:10.1103/PhysRevD.79.036002
  [arXiv:0811.2955 [hep-ph]].
  %%CITATION = doi:10.1103/PhysRevD.79.036002;%%
  %56 citations counted in INSPIRE as of 13 Jan 2016



%\cite{Cao:2010mp}
\bibitem{Cao:2010mp}
  Q.~H.~Cao, S.~Khalil, E.~Ma and H.~Okada,
  %``Observable $T_7$ Lepton Flavor Symmetry at the Large Hadron Collider,''
  Phys.\ Rev.\ Lett.\  {\bf 106}, 131801 (2011)
  doi:10.1103/PhysRevLett.106.131801
  [arXiv:1009.5415 [hep-ph]].
  %%CITATION = doi:10.1103/PhysRevLett.106.131801;%%
  %27 citations counted in INSPIRE as of 13 Jan 2016



%\cite{Luhn:2012bc}
\bibitem{Luhn:2012bc}
  C.~Luhn, K.~M.~Parattu and A.~Wingerter,
  %``A Minimal Model of Neutrino Flavor,''
  JHEP {\bf 1212}, 096 (2012)
  doi:10.1007/JHEP12(2012)096
  [arXiv:1210.1197 [hep-ph]].
  %%CITATION = doi:10.1007/JHEP12(2012)096;%%
  %17 citations counted in INSPIRE as of 13 Jan 2016



%\cite{Kajiyama:2013lja}
\bibitem{Kajiyama:2013lja}
  Y.~Kajiyama, H.~Okada and K.~Yagyu,
  %``$T_7$ Flavor Model in Three Loop Seesaw and Higgs Phenomenology,''
  JHEP {\bf 1310}, 196 (2013)
  doi:10.1007/JHEP10(2013)196
  [arXiv:1307.0480 [hep-ph]].
  %%CITATION = doi:10.1007/JHEP10(2013)196;%%
  %33 citations counted in INSPIRE as of 13 Jan 2016



%\cite{Bonilla:2014xla}
\bibitem{Bonilla:2014xla}
  C.~Bonilla, S.~Morisi, E.~Peinado and J.~W.~F.~Valle,
  %``Relating quarks and leptons with the $T_7$ flavour group,''
  Phys.\ Lett.\ B {\bf 742}, 99 (2015)
  doi:10.1016/j.physletb.2015.01.017
  [arXiv:1411.4883 [hep-ph]].
  %%CITATION = doi:10.1016/j.physletb.2015.01.017;%%
  %8 citations counted in INSPIRE as of 13 Jan 2016



%\cite{Vien:2014gza}
\bibitem{Vien:2014gza}
  V.~V.~Vien and H.~N.~Long,
  %``The $T_7$ flavor symmetry in 3-3-1 model with neutral leptons,''
  JHEP {\bf 1404}, 133 (2014)
  doi:10.1007/JHEP04(2014)133
  [arXiv:1402.1256 [hep-ph]].
  %%CITATION = doi:10.1007/JHEP04(2014)133;%%
  %15 citations counted in INSPIRE as of 13 Jan 2016



%\cite{Vien:2015koa}
\bibitem{Vien:2015koa}
  V.~V.~Vien,
  %``$T_7$ flavor symmetry scheme for understanding neutrino mass and mixing in 3-3-1 model with neutral leptons,''
  Mod.\ Phys.\ Lett.\ A {\bf 29}, 28 (2014)
  doi:10.1142/S0217732314501399
  [arXiv:1508.02585 [hep-ph]].
  %%CITATION = doi:10.1142/S0217732314501399;%%



%\cite{Hernandez:2015cra}
\bibitem{Hernandez:2015cra}
  A.~E.~C.~Hernández and R.~Martinez,
  %``Fermion mass and mixing pattern in a minimal T7 flavor 331 model,''
  arXiv:1501.07261 [hep-ph].
  %%CITATION = ARXIV:1501.07261;%%
  %12 citations counted in INSPIRE as of 13 Jan 2016



%\cite{Arbelaez:2015toa}
\bibitem{Arbelaez:2015toa}
  C.~Arbeláez, A.~E.~Cárcamo Hernández, S.~Kovalenko and I.~Schmidt,
  %``Adjoint $SU(5)$ GUT model with $T_{7}$ flavor symmetry,''
  Phys.\ Rev.\ D {\bf 92}, no. 11, 115015 (2015)
  doi:10.1103/PhysRevD.92.115015
  [arXiv:1507.03852 [hep-ph]].
  %%CITATION = doi:10.1103/PhysRevD.92.115015;%%
  %3 citations counted in INSPIRE as of 13 Jan 2016



%\cite{Ding:2011qt}
\bibitem{Ding:2011qt}
  G.~J.~Ding,
  %``Tri-Bimaximal Neutrino Mixing and the $T_{13}$ Flavor Symmetry,''
  Nucl.\ Phys.\ B {\bf 853}, 635 (2011)
  doi:10.1016/j.nuclphysb.2011.08.012
  [arXiv:1105.5879 [hep-ph]].
  %%CITATION = doi:10.1016/j.nuclphysb.2011.08.012;%%
  %8 citations counted in INSPIRE as of 13 Jan 2016



%\cite{Hartmann:2011dn}
\bibitem{Hartmann:2011dn}
  C.~Hartmann,
  %``The Frobenius group T13 and the canonical see-saw mechanism applied to neutrino mixing,''
  Phys.\ Rev.\ D {\bf 85}, 013012 (2012)
  doi:10.1103/PhysRevD.85.013012
  [arXiv:1109.5143 [hep-ph]].
  %%CITATION = doi:10.1103/PhysRevD.85.013012;%%
  %2 citations counted in INSPIRE as of 13 Jan 2016



%\cite{Hartmann:2011pq}
\bibitem{Hartmann:2011pq}
  C.~Hartmann and A.~Zee,
  %``Neutrino Mixing and the Frobenius Group T13,''
  Nucl.\ Phys.\ B {\bf 853}, 105 (2011)
  doi:10.1016/j.nuclphysb.2011.07.023
  [arXiv:1106.0333 [hep-ph]].
  %%CITATION = doi:10.1016/j.nuclphysb.2011.07.023;%%
  %6 citations counted in INSPIRE as of 13 Jan 2016



%\cite{Kajiyama:2010sb}
\bibitem{Kajiyama:2010sb}
  Y.~Kajiyama and H.~Okada,
  %``T(13) Flavor Symmetry and Decaying Dark Matter,''
  Nucl.\ Phys.\ B {\bf 848}, 303 (2011)
  doi:10.1016/j.nuclphysb.2011.02.020
  [arXiv:1011.5753 [hep-ph]].
  %%CITATION = doi:10.1016/j.nuclphysb.2011.02.020;%%
  %28 citations counted in INSPIRE as of 13 Jan 2016



%\cite{Aranda:2000tm}
\bibitem{Aranda:2000tm}
  A.~Aranda, C.~D.~Carone and R.~F.~Lebed,
  %``Maximal neutrino mixing from a minimal flavor symmetry,''
  Phys.\ Rev.\ D {\bf 62}, 016009 (2000)
  doi:10.1103/PhysRevD.62.016009
  [hep-ph/0002044].
  %%CITATION = doi:10.1103/PhysRevD.62.016009;%%
  %107 citations counted in INSPIRE as of 13 Jan 2016



%\cite{Aranda:2007dp}
\bibitem{Aranda:2007dp}
  A.~Aranda,
  %``Neutrino mixing from the double tetrahedral group T-prime,''
  Phys.\ Rev.\ D {\bf 76}, 111301 (2007)
  doi:10.1103/PhysRevD.76.111301
  [arXiv:0707.3661 [hep-ph]].
  %%CITATION = doi:10.1103/PhysRevD.76.111301;%%
  %68 citations counted in INSPIRE as of 13 Jan 2016



%\cite{Chen:2007afa}
\bibitem{Chen:2007afa}
  M.~C.~Chen and K.~T.~Mahanthappa,
  %``CKM and Tri-bimaximal MNS Matrices in a $SU(5) \times ^{(d)}T$ Model,''
  Phys.\ Lett.\ B {\bf 652}, 34 (2007)
  doi:10.1016/j.physletb.2007.06.064
  [arXiv:0705.0714 [hep-ph]].
  %%CITATION = doi:10.1016/j.physletb.2007.06.064;%%
  %197 citations counted in INSPIRE as of 13 Jan 2016



%\cite{Frampton:2008bz}
\bibitem{Frampton:2008bz}
  P.~H.~Frampton, T.~W.~Kephart and S.~Matsuzaki,
  %``Simplified Renormalizable T-prime Model for Tribimaximal Mixing and Cabibbo Angle,''
  Phys.\ Rev.\ D {\bf 78}, 073004 (2008)
  doi:10.1103/PhysRevD.78.073004
  [arXiv:0807.4713 [hep-ph]].
  %%CITATION = doi:10.1103/PhysRevD.78.073004;%%
  %51 citations counted in INSPIRE as of 13 Jan 2016



%\cite{Eby:2011ph}
\bibitem{Eby:2011ph}
  D.~A.~Eby, P.~H.~Frampton, X.~G.~He and T.~W.~Kephart,
  %``Quartification with T' Flavor,''
  Phys.\ Rev.\ D {\bf 84}, 037302 (2011)
  doi:10.1103/PhysRevD.84.037302
  [arXiv:1103.5737 [hep-ph]].
  %%CITATION = doi:10.1103/PhysRevD.84.037302;%%
  %6 citations counted in INSPIRE as of 13 Jan 2016



%\cite{Frampton:2013lva}
\bibitem{Frampton:2013lva}
  P.~H.~Frampton, C.~M.~Ho and T.~W.~Kephart,
  %``Heterotic discrete flavor model,''
  Phys.\ Rev.\ D {\bf 89}, no. 2, 027701 (2014)
  doi:10.1103/PhysRevD.89.027701
  [arXiv:1305.4402 [hep-ph]].
  %%CITATION = doi:10.1103/PhysRevD.89.027701;%%
  %4 citations counted in INSPIRE as of 13 Jan 2016



%\cite{Varzielas:2012nn}
\bibitem{Varzielas:2012nn}
  I.~de Medeiros Varzielas, D.~Emmanuel-Costa and P.~Leser,
  %``Geometrical CP Violation from Non-Renormalisable Scalar Potentials,''
  Phys.\ Lett.\ B {\bf 716}, 193 (2012)
  doi:10.1016/j.physletb.2012.08.008
  [arXiv:1204.3633 [hep-ph]].
  %%CITATION = doi:10.1016/j.physletb.2012.08.008;%%
  %37 citations counted in INSPIRE as of 13 Jan 2016



%\cite{Bhattacharyya:2012pi}
\bibitem{Bhattacharyya:2012pi}
  G.~Bhattacharyya, I.~de Medeiros Varzielas and P.~Leser,
  %``A common origin of fermion mixing and geometrical CP violation, and its test through Higgs physics at the LHC,''
  Phys.\ Rev.\ Lett.\  {\bf 109}, 241603 (2012)
  doi:10.1103/PhysRevLett.109.241603
  [arXiv:1210.0545 [hep-ph]].
  %%CITATION = doi:10.1103/PhysRevLett.109.241603;%%
  %49 citations counted in INSPIRE as of 13 Jan 2016



%\cite{Ma:2013xqa}
\bibitem{Ma:2013xqa}
  E.~Ma,
  %``Neutrino Mixing and Geometric CP Violation with Delta(27) Symmetry,''
  Phys.\ Lett.\ B {\bf 723}, 161 (2013)
  doi:10.1016/j.physletb.2013.05.011
  [arXiv:1304.1603 [hep-ph]].
  %%CITATION = doi:10.1016/j.physletb.2013.05.011;%%
  %17 citations counted in INSPIRE as of 13 Jan 2016



%\cite{Nishi:2013jqa}
\bibitem{Nishi:2013jqa}
  C.~C.~Nishi,
  %``Generalized $CP$ symmetries in $\Delta(27)$ flavor models,''
  Phys.\ Rev.\ D {\bf 88}, no. 3, 033010 (2013)
  doi:10.1103/PhysRevD.88.033010
  [arXiv:1306.0877 [hep-ph]].
  %%CITATION = doi:10.1103/PhysRevD.88.033010;%%
  %27 citations counted in INSPIRE as of 13 Jan 2016



%\cite{Varzielas:2013sla}
\bibitem{Varzielas:2013sla}
  I.~de Medeiros Varzielas and D.~Pidt,
  %``Towards realistic models of quark masses with geometrical CP violation,''
  J.\ Phys.\ G {\bf 41}, 025004 (2014)
  doi:10.1088/0954-3899/41/2/025004
  [arXiv:1307.0711 [hep-ph]].
  %%CITATION = doi:10.1088/0954-3899/41/2/025004;%%
  %24 citations counted in INSPIRE as of 13 Jan 2016



%\cite{Aranda:2013gga}
\bibitem{Aranda:2013gga}
  A.~Aranda, C.~Bonilla, S.~Morisi, E.~Peinado and J.~W.~F.~Valle,
  %``Dirac neutrinos from flavor symmetry,''
  Phys.\ Rev.\ D {\bf 89}, no. 3, 033001 (2014)
  doi:10.1103/PhysRevD.89.033001
  [arXiv:1307.3553 [hep-ph]].
  %%CITATION = doi:10.1103/PhysRevD.89.033001;%%
  %11 citations counted in INSPIRE as of 13 Jan 2016



%\cite{Varzielas:2015aua}
\bibitem{Varzielas:2015aua}
  I.~de Medeiros Varzielas,
  %``$\Delta(27)$ family symmetry and neutrino mixing,''
  JHEP {\bf 1508}, 157 (2015)
  doi:10.1007/JHEP08(2015)157
  [arXiv:1507.00338 [hep-ph]].
  %%CITATION = doi:10.1007/JHEP08(2015)157;%%
  %3 citations counted in INSPIRE as of 13 Jan 2016



%\cite{Chen:2015jta}
\bibitem{Chen:2015jta}
  P.~Chen, G.~J.~Ding, A.~D.~Rojas, C.~A.~Vaquera-Araujo and J.~W.~F.~Valle,
  %``Warped flavor symmetry predictions for neutrino physics,''
  JHEP {\bf 1601}, 007 (2016)
  doi:10.1007/JHEP01(2016)007
  [arXiv:1509.06683 [hep-ph]].
  %%CITATION = doi:10.1007/JHEP01(2016)007;%%
  %1 citations counted in INSPIRE as of 13 Jan 2016



%\cite{Georgi:1978bv}
\bibitem{Georgi:1978bv}
  H.~Georgi and A.~Pais,
  %``Generalization of Gim: Horizontal and Vertical Flavor Mixing,''
  Phys.\ Rev.\ D {\bf 19}, 2746 (1979).
  doi:10.1103/PhysRevD.19.2746
  %%CITATION = doi:10.1103/PhysRevD.19.2746;%%
  %54 citations counted in INSPIRE as of 13 Jan 2016



%\cite{Valle:1983dk}
\bibitem{Valle:1983dk}
  J.~W.~F.~Valle and M.~Singer,
  %``Lepton Number Violation With Quasi Dirac Neutrinos,''
  Phys.\ Rev.\ D {\bf 28}, 540 (1983).
  doi:10.1103/PhysRevD.28.540
  %%CITATION = doi:10.1103/PhysRevD.28.540;%%
  %106 citations counted in INSPIRE as of 13 Jan 2016



%\cite{Pisano:1991ee}
\bibitem{Pisano:1991ee}
  F.~Pisano and V.~Pleitez,
  %``An SU(3) x U(1) model for electroweak interactions,''
  Phys.\ Rev.\ D {\bf 46}, 410 (1992)
  doi:10.1103/PhysRevD.46.410
  [hep-ph/9206242].
  %%CITATION = doi:10.1103/PhysRevD.46.410;%%
  %522 citations counted in INSPIRE as of 13 Jan 2016



%\cite{Foot:1992rh}
\bibitem{Foot:1992rh}
  R.~Foot, O.~F.~Hernandez, F.~Pisano and V.~Pleitez,
  %``Lepton masses in an SU(3)-L x U(1)-N gauge model,''
  Phys.\ Rev.\ D {\bf 47}, 4158 (1993)
  doi:10.1103/PhysRevD.47.4158
  [hep-ph/9207264].
  %%CITATION = doi:10.1103/PhysRevD.47.4158;%%
  %264 citations counted in INSPIRE as of 13 Jan 2016



%\cite{Frampton:1992wt}
\bibitem{Frampton:1992wt}
  P.~H.~Frampton,
  %``Chiral dilepton model and the flavor question,''
  Phys.\ Rev.\ Lett.\  {\bf 69}, 2889 (1992).
  doi:10.1103/PhysRevLett.69.2889
  %%CITATION = doi:10.1103/PhysRevLett.69.2889;%%
  %527 citations counted in INSPIRE as of 13 Jan 2016



%\cite{Ng:1992st}
\bibitem{Ng:1992st}
  D.~Ng,
  %``The Electroweak theory of SU(3) x U(1),''
  Phys.\ Rev.\ D {\bf 49}, 4805 (1994)
  doi:10.1103/PhysRevD.49.4805
  [hep-ph/9212284].
  %%CITATION = doi:10.1103/PhysRevD.49.4805;%%
  %155 citations counted in INSPIRE as of 13 Jan 2016



%\cite{Duong:1993zn}
\bibitem{Duong:1993zn}
  T.~V.~Duong and E.~Ma,
  %``Supersymmetric SU(3) x U(1) gauge models: Higgs structure at the electroweak energy scale,''
  Phys.\ Lett.\ B {\bf 316}, 307 (1993)
  doi:10.1016/0370-2693(93)90329-G
  [hep-ph/9306264].
  %%CITATION = doi:10.1016/0370-2693(93)90329-G;%%
  %65 citations counted in INSPIRE as of 13 Jan 2016



%\cite{Hoang:1996gi}
\bibitem{Hoang:1996gi}
  H.~N.~Long,
  %``SU(3)-L x U(1)-N model for right-handed neutrino neutral currents,''
  Phys.\ Rev.\ D {\bf 54}, 4691 (1996)
  doi:10.1103/PhysRevD.54.4691
  [hep-ph/9607439].
  %%CITATION = doi:10.1103/PhysRevD.54.4691;%%
  %159 citations counted in INSPIRE as of 13 Jan 2016



%\cite{Hoang:1995vq}
\bibitem{Hoang:1995vq}
  H.~N.~Long,
  %``The 331 model with right handed neutrinos,''
  Phys.\ Rev.\ D {\bf 53}, 437 (1996)
  doi:10.1103/PhysRevD.53.437
  [hep-ph/9504274].
  %%CITATION = doi:10.1103/PhysRevD.53.437;%%
  %173 citations counted in INSPIRE as of 13 Jan 2016



%\cite{Foot:1994ym}
\bibitem{Foot:1994ym}
  R.~Foot, H.~N.~Long and T.~A.~Tran,
  %``SU(3)-L x U(1)-N and SU(4)-L x U(1)-N gauge models with right-handed neutrinos,''
  Phys.\ Rev.\ D {\bf 50}, R 34 (1994)
  doi:10.1103/PhysRevD.50.R34
  [hep-ph/9402243].
  %%CITATION = doi:10.1103/PhysRevD.50.R34;%%
  %236 citations counted in INSPIRE as of 13 Jan 2016



%\cite{Diaz:2003dk}
\bibitem{Diaz:2003dk}
  R.~A.~Diaz, R.~Martinez and F.~Ochoa,
  %``The Scalar sector of the SU(3)(c) x SU(3)(L) x U(1)(X) model,''
  Phys.\ Rev.\ D {\bf 69}, 095009 (2004)
  doi:10.1103/PhysRevD.69.095009
  [hep-ph/0309280].
  %%CITATION = doi:10.1103/PhysRevD.69.095009;%%
  %67 citations counted in INSPIRE as of 13 Jan 2016



%\cite{Diaz:2004fs}
\bibitem{Diaz:2004fs}
  R.~A.~Diaz, R.~Martinez and F.~Ochoa,
  %``SU(3)(c) x SU(3)(L) x U(1)(X) models for beta arbitrary and families with mirror fermions,''
  Phys.\ Rev.\ D {\bf 72}, 035018 (2005)
  doi:10.1103/PhysRevD.72.035018
  [hep-ph/0411263].
  %%CITATION = doi:10.1103/PhysRevD.72.035018;%%
  %62 citations counted in INSPIRE as of 13 Jan 2016



%\cite{Dias:2004dc}
\bibitem{Dias:2004dc}
  A.~G.~Dias, R.~Martinez and V.~Pleitez,
  %``Concerning the Landau pole in 3-3-1 models,''
  Eur.\ Phys.\ J.\ C {\bf 39}, 101 (2005)
  doi:10.1140/epjc/s2004-02083-0
  [hep-ph/0407141].
  %%CITATION = doi:10.1140/epjc/s2004-02083-0;%%
  %66 citations counted in INSPIRE as of 13 Jan 2016



%\cite{Ochoa:2005ch}
\bibitem{Ochoa:2005ch}
  F.~Ochoa and R.~Martinez,
  %``Family dependence in SU(3)(c) x SU(3)(L) x U(1)(X) models,''
  Phys.\ Rev.\ D {\bf 72}, 035010 (2005)
  doi:10.1103/PhysRevD.72.035010
  [hep-ph/0505027].
  %%CITATION = doi:10.1103/PhysRevD.72.035010;%%
  %30 citations counted in INSPIRE as of 13 Jan 2016



%\cite{CarcamoHernandez:2005ka}
\bibitem{CarcamoHernandez:2005ka}
  A.~E.~Carcamo Hernandez, R.~Martinez and F.~Ochoa,
  %``Z and Z' decays with and without FCNC in 331 models,''
  Phys.\ Rev.\ D {\bf 73}, 035007 (2006)
  doi:10.1103/PhysRevD.73.035007
  [hep-ph/0510421].
  %%CITATION = doi:10.1103/PhysRevD.73.035007;%%
  %57 citations counted in INSPIRE as of 13 Jan 2016



%\cite{Alvarado:2012xi}
\bibitem{Alvarado:2012xi}
  C.~Alvarado, R.~Martinez and F.~Ochoa,
  %``Quark mass hierarchy in 3-3-1 models,''
  Phys.\ Rev.\ D {\bf 86}, 025027 (2012)
  doi:10.1103/PhysRevD.86.025027
  [arXiv:1207.0014 [hep-ph]].
  %%CITATION = doi:10.1103/PhysRevD.86.025027;%%
  %13 citations counted in INSPIRE as of 13 Jan 2016



%\cite{Catano:2012kw}
\bibitem{Catano:2012kw}
  M.~E.~Catano, R.~Martinez and F.~Ochoa,
  %``Neutrino masses in a 331 model with right-handed neutrinos without doubly charged Higgs bosons via inverse and double seesaw mechanisms,''
  Phys.\ Rev.\ D {\bf 86}, 073015 (2012)
  doi:10.1103/PhysRevD.86.073015
  [arXiv:1206.1966 [hep-ph]].
  %%CITATION = doi:10.1103/PhysRevD.86.073015;%%
  %18 citations counted in INSPIRE as of 13 Jan 2016



%\cite{Boucenna:2014ela}
\bibitem{Boucenna:2014ela}
  S.~M.~Boucenna, S.~Morisi and J.~W.~F.~Valle,
  %``Radiative neutrino mass in 3-3-1 scheme,''
  Phys.\ Rev.\ D {\bf 90}, no. 1, 013005 (2014)
  doi:10.1103/PhysRevD.90.013005
  [arXiv:1405.2332 [hep-ph]].
  %%CITATION = doi:10.1103/PhysRevD.90.013005;%%
  %15 citations counted in INSPIRE as of 13 Jan 2016



%\cite{Boucenna:2014dia}
\bibitem{Boucenna:2014dia}
  S.~M.~Boucenna, R.~M.~Fonseca, F.~Gonzalez-Canales and J.~W.~F.~Valle,
  %``Small neutrino masses and gauge coupling unification,''
  Phys.\ Rev.\ D {\bf 91}, no. 3, 031702 (2015)
  doi:10.1103/PhysRevD.91.031702
  [arXiv:1411.0566 [hep-ph]].
  %%CITATION = doi:10.1103/PhysRevD.91.031702;%%
  %10 citations counted in INSPIRE as of 13 Jan 2016



%\cite{Phong:2014ofa}
\bibitem{Phong:2014ofa}
  V.~Q.~Phong, H.~N.~Long, V.~T.~Van and L.~H.~Minh,
  %``Electroweak phase transition in the economical 3-3-1 model,''
  Eur.\ Phys.\ J.\ C {\bf 75}, no. 7, 342 (2015)
  doi:10.1140/epjc/s10052-015-3550-2
  [arXiv:1409.0750 [hep-ph]].
  %%CITATION = doi:10.1140/epjc/s10052-015-3550-2;%%
  %3 citations counted in INSPIRE as of 13 Jan 2016



%\cite{Boucenna:2015zwa}
\bibitem{Boucenna:2015zwa}
  S.~M.~Boucenna, J.~W.~F.~Valle and A.~Vicente,
  %``Predicting charged lepton flavor violation from 3-3-1 gauge symmetry,''
  Phys.\ Rev.\ D {\bf 92}, no. 5, 053001 (2015)
  doi:10.1103/PhysRevD.92.053001
  [arXiv:1502.07546 [hep-ph]].
  %%CITATION = doi:10.1103/PhysRevD.92.053001;%%
  %6 citations counted in INSPIRE as of 13 Jan 2016



%\cite{DeConto:2015eia}
\bibitem{DeConto:2015eia}
  G.~De Conto, A.~C.~B.~Machado and V.~Pleitez,
  %``Minimal 3-3-1 model with a spectator sextet,''
  Phys.\ Rev.\ D {\bf 92}, no. 7, 075031 (2015)
  doi:10.1103/PhysRevD.92.075031
  [arXiv:1505.01343 [hep-ph]].
  %%CITATION = doi:10.1103/PhysRevD.92.075031;%%
  %1 citations counted in INSPIRE as of 13 Jan 2016



%\cite{Correia:2015tra}
\bibitem{Correia:2015tra}
  F.~C.~Correia and V.~Pleitez,
  %``Neutral meson mixing induced by box diagrams in the 3-3-1 model with heavy leptons,''
  Phys.\ Rev.\ D {\bf 92}, 113006 (2015)
  doi:10.1103/PhysRevD.92.113006
  [arXiv:1508.07319 [hep-ph]].
  %%CITATION = doi:10.1103/PhysRevD.92.113006;%%



%\cite{Okada:2015bxa}
\bibitem{Okada:2015bxa}
  H.~Okada, N.~Okada and Y.~Orikasa,
  %``Radiative Seesaw in Minimal 3-3-1 Model,''
  Phys. \ Rev.\ D \ 93, 073006 (2016)
  doi:10.1103/PhysRevD.93.073006
  arXiv:1504.01204 [hep-ph].
  %%CITATION = ARXIV:1504.01204;%%
  %11 citations counted in INSPIRE as of 13 Jan 2016



%\cite{Long:2015gca}
\bibitem{Long:2015gca}
  H.~N.~Long,
  %``Michel parameter in 3-3-1 model with three lepton singlets,''
  arXiv:1504.06908 [hep-ph].
  %%CITATION = ARXIV:1504.06908;%%



%\cite{Long:2015qza}
\bibitem{Long:2015qza}
  H.~N.~Long, \textbf{Physics International, vol. 7, No 1 (2016)  1, }
  %``Early Universe in the SU(3)_L X U(1)_X electroweak models,''
  arXiv:1501.01852 [hep-ph].
  %%CITATION = ARXIV:1501.01852;%%



%\cite{Binh:2015cba}
\bibitem{Binh:2015cba}
  D.~T.~Binh, D.~T.~Huong and H.~N.~Long,
  %``The muon anomalous magnetic moment in the supersymmetric economical 3-3-1 model,''
  Zh.\ Eksp.\ Teor.\ Fiz.\  {\bf 148}, 1115 (2015)
  doi:10.7868/S004445101512007X
  [arXiv:1504.03510 [hep-ph]].
  %%CITATION = doi:10.7868/S004445101512007X;%%



%\cite{Hue:2015fbb}
\bibitem{Hue:2015fbb}
  L.~T.~Hue, H.~N.~Long, T.~T.~Thuc and N.~T.~Phong,
  %``Lepton flavor violating decays of Standard-Model-like Higgs in 3-3-1 model with neutral lepton,''
  Nucl. \ Phys. \ {\bf  B  907}, (2016) 37.
  doi:10.1016/j.nuclphysb.2016.03.034
  arXiv:1512.03266 [hep-ph].
  %%CITATION = ARXIV:1512.03266;%%



%\cite{Pal:1994ba}
\bibitem{Pal:1994ba}
  P.~B.~Pal,
  %``The Strong CP question in SU(3)(C) x SU(3)(L) x U(1)(N) models,''
  Phys.\ Rev.\ D {\bf 52}, 1659 (1995)
  doi:10.1103/PhysRevD.52.1659
  [hep-ph/9411406].
  %%CITATION = doi:10.1103/PhysRevD.52.1659;%%
  %78 citations counted in INSPIRE as of 13 Jan 2016



%\cite{Mizukoshi:2010ky}
\bibitem{Mizukoshi:2010ky}
  J.~K.~Mizukoshi, C.~A.~de S.Pires, F.~S.~Queiroz and P.~S.~Rodrigues da Silva,
  %``WIMPs in a 3-3-1 model with heavy Sterile neutrinos,''
  Phys.\ Rev.\ D {\bf 83}, 065024 (2011)
  doi:10.1103/PhysRevD.83.065024
  [arXiv:1010.4097 [hep-ph]].
  %%CITATION = doi:10.1103/PhysRevD.83.065024;%%
  %41 citations counted in INSPIRE as of 13 Jan 2016



%\cite{Boucenna:2015pav}
\bibitem{Boucenna:2015pav}
  S.~M.~Boucenna, S.~Morisi and A.~Vicente,
  %``The LHC diphoton resonance from gauge symmetry,''
  arXiv:1512.06878 [hep-ph].
  %%CITATION = ARXIV:1512.06878;%%
  %48 citations counted in INSPIRE as of 13 Jan 2016



%\cite{Hernandez:2015ywg}
\bibitem{Hernandez:2015ywg}
  A.~E.~C.~Hernández and I.~Nisandzic,
  %``LHC diphoton 750 GeV resonance as an indication of $SU(3)_c\times SU(3)_L\times U(1)_X$ gauge symmetry,''
  arXiv:1512.07165 [hep-ph].
  %%CITATION = ARXIV:1512.07165;%%
  %50 citations counted in INSPIRE as of 13 Jan 2016



%\cite{Dong:2015dxw}
\bibitem{Dong:2015dxw}
  P.~V.~Dong and N.~T.~K.~Ngan,
  %``Phenomenology of the simple 3-3-1 model with inert scalars,''
  arXiv:1512.09073 [hep-ph].
  %%CITATION = ARXIV:1512.09073;%%
  %6 citations counted in INSPIRE as of 13 Jan 2016



%\cite{331Pisano:1992}
\bibitem{331Pisano:1992}
F. ~Pisano and V. ~Pleitez, Phys. Rev. D {\bf 46}, 410 (1992),
  %doi:10.1103/PhysRevD.46.410
  [arXiv: hep-ph/9206242].






%\cite{331Frampton:1992}
\bibitem{331Frampton:1992}
 P. ~H. ~Frampton, Phys. Rev. Lett. {\bf 69} (1992) 2889.




%\cite{331Foot:1993}
\bibitem{331Foot:1993}
R. ~Foot et al. Phys. Rev. D {\bf 47} (1993) 4158.




%\cite{Chang:2006aa}
\bibitem{Chang:2006aa}
  D.~Chang and H.~N.~Long,
  %``Interesting radiative patterns of neutrino mass in an SU(3)(C) x SU(3)(L) x U(1)(X) model with right-handed neutrinos,''
  Phys.\ Rev.\ D {\bf 73}, 053006 (2006)
  doi:10.1103/PhysRevD.73.053006
  [hep-ph/0603098].
  %%CITATION = doi:10.1103/PhysRevD.73.053006;%%
  %56 citations counted in INSPIRE as of 13 Jan 2016



%\cite{HarrisonPS:2002}
\bibitem{HarrisonPS:2002}
  P.F. Harrison, D.H. Perkins, W.G. Scott,
  %``Tri-Bimaximal Mixing and the Neutrino Oscillation Data'',
  Phys.\ Lett.\ B 530, 167 (2002)
  doi: 10.1016/S0370-2693(02)01336-9
  [hep-ph/0202074].

%\cite{GonzalezGarcia:2014sz}
\bibitem{GonzalezGarcia:2014sz}
  M.~C.~Gonzalez-Garcia, M.~Maltoni and T.~Schwetz,
  %``Updated fit to three neutrino mixing: status of leptonic CP violation,''
  JHEP {\bf1411}, 052 (2014)
  doi:10.1007/JHEP11(2014)052
  [arXiv:1409.5439 [hep-ph]].


%\cite{Conrad:2013dma}
\bibitem{Conrad:2013dma}
  J.~M.~Conrad,
  %``Neutrino experiments and the Large Hadron Collider: friends across 14 orders of magnitude,''
  Phys.\ Scripta T {\bf 158}, 014012 (2013)
  doi:10.1088/0031-8949/2013/T158/014012
  [arXiv:1310.0108 [hep-ex]].
  %%CITATION = doi:10.1088/0031-8949/2013/T158/014012;%%
  %1 citations counted in INSPIRE as of 13 Jan 2016



%\cite{Parke:2013pna}
\bibitem{Parke:2013pna}
  S.~Parke,
  %``Neutrinos: Theory and Phenomenology,''
  Phys.\ Scripta T {\bf 158}, 014013 (2013)
  doi:10.1088/0031-8949/2013/T158/014013
  [arXiv:1310.5992 [hep-ph]].
  %%CITATION = doi:10.1088/0031-8949/2013/T158/014013;%%
  %12 citations counted in INSPIRE as of 13 Jan 2016



%\cite{Tegmark:2003ud}
\bibitem{Tegmark:2003ud}
  M.~Tegmark {\it et al.} [SDSS Collaboration],
  %``Cosmological parameters from SDSS and WMAP,''
  Phys.\ Rev.\ D {\bf 69}, 103501 (2004)
  doi:10.1103/PhysRevD.69.103501
  [astro-ph/0310723].
  %%CITATION = doi:10.1103/PhysRevD.69.103501;%%
  %2463 citations counted in INSPIRE as of 13 Jan 2016



%\cite{Weiler:2013rta}
\bibitem{Weiler:2013rta}
  T.~J.~Weiler,
  %``Oscillation and Mixing Among the Three Neutrino Flavors,''
  arXiv:1308.1715 [hep-ph].
  %%CITATION = ARXIV:1308.1715;%%
  %6 citations counted in INSPIRE as of 13 Jan 2016



%\cite{Auger:2012ar}
\bibitem{Auger:2012ar}
  M.~Auger {\it et al.} [EXO-200 Collaboration],
  %``Search for Neutrinoless Double-Beta Decay in $^{136}$Xe with EXO-200,''
  Phys.\ Rev.\ Lett.\  {\bf 109}, 032505 (2012)
  doi:10.1103/PhysRevLett.109.032505
  [arXiv:1205.5608 [hep-ex]].
  %%CITATION = doi:10.1103/PhysRevLett.109.032505;%%
  %317 citations counted in INSPIRE as of 13 Jan 2016



%\cite{Abt:2004yk}
\bibitem{Abt:2004yk}
  I.~Abt {\it et al.},
  %``A New $Ge^{76}$ Double Beta Decay Experiment at LNGS: Letter of Intent,''
  hep-ex/0404039.
  %%CITATION = HEP-EX/0404039;%%
  %212 citations counted in INSPIRE as of 13 Jan 2016



%\cite{Ackermann:2012xja}
\bibitem{Ackermann:2012xja}
  K.~H.~Ackermann {\it et al.} [GERDA Collaboration],
  %``The GERDA experiment for the search of $0\nu\beta\beta$  decay in $^{76}$Ge,''
  Eur.\ Phys.\ J.\ C {\bf 73}, no. 3, 2330 (2013)
  doi:10.1140/epjc/s10052-013-2330-0
  [arXiv:1212.4067 [physics.ins-det]].
  %%CITATION = doi:10.1140/epjc/s10052-013-2330-0;%%
  %100 citations counted in INSPIRE as of 13 Jan 2016



%\cite{Alessandria:2011rc}
\bibitem{Alessandria:2011rc}
  F.~Alessandria {\it et al.},
  %``Sensitivity of CUORE to Neutrinoless Double-Beta Decay,''
  arXiv:1109.0494 [nucl-ex].
  %%CITATION = ARXIV:1109.0494;%%
  %55 citations counted in INSPIRE as of 13 Jan 2016



%\cite{KamLANDZen:2012aa}
\bibitem{KamLANDZen:2012aa}
  A.~Gando {\it et al.} [KamLAND-Zen Collaboration],
  %``Measurement of the double-\beta decay half-life of ^{136}Xe with the KamLAND-Zen experiment,''
  Phys.\ Rev.\ C {\bf 85}, 045504 (2012)
  doi:10.1103/PhysRevC.85.045504
  [arXiv:1201.4664 [hep-ex]].
  %%CITATION = doi:10.1103/PhysRevC.85.045504;%%
  %146 citations counted in INSPIRE as of 13 Jan 2016



%\cite{Albert:2014fya}
\bibitem{Albert:2014fya}
  J.~B.~Albert {\it et al.} [EXO-200 Collaboration],
  %``Search for Majoron-emitting modes of double-beta decay of $^{136}$Xe with EXO-200,''
  Phys.\ Rev.\ D {\bf 90}, no. 9, 092004 (2014)
  doi:10.1103/PhysRevD.90.092004
  [arXiv:1409.6829 [hep-ex]].
  %%CITATION = doi:10.1103/PhysRevD.90.092004;%%
  %10 citations counted in INSPIRE as of 13 Jan 2016



%\cite{Guiseppe:2011me}
\bibitem{Guiseppe:2011me}
  C.~E.~Aalseth {\it et al.} [Majorana Collaboration],
  %``The Majorana Experiment,''
  Nucl.\ Phys.\ Proc.\ Suppl.\  {\bf 217}, 44 (2011)
  doi:10.1016/j.nuclphysbps.2011.04.063
  [arXiv:1101.0119 [nucl-ex]].
  %%CITATION = doi:10.1016/j.nuclphysbps.2011.04.063;%%
  %21 citations counted in INSPIRE as of 13 Jan 2016



%\cite{Bilenky:2014uka}
\bibitem{Bilenky:2014uka}
  S.~M.~Bilenky and C.~Giunti,
  %``Neutrinoless Double-Beta Decay: a Probe of Physics Beyond the Standard Model,''
  Int.\ J.\ Mod.\ Phys.\ A {\bf 30}, no. 04n05, 1530001 (2015)
  doi:10.1142/S0217751X1530001X
  [arXiv:1411.4791 [hep-ph]].
  %%CITATION = doi:10.1142/S0217751X1530001X;%%
  %38 citations counted in INSPIRE as of 13 Jan 2016

%\cite{Okada:2016whh}
\bibitem{Okada:2016whh}
  H.~Okada, N.~Okada, Y.~Orikasa and K.~Yagyu,
  %``Higgs phenomenology in the minimal SU(3)$_L$×U(1)$_X$ model,''
  Phys.\ Rev.\ D {\bf 94}, no. 1, 015002 (2016)
  doi:10.1103/PhysRevD.94.015002
  [arXiv:1604.01948 [hep-ph]].
  %%CITATION = doi:10.1103/PhysRevD.94.015002;%%
  %4 citations counted in INSPIRE as of 01 Sep 2016

%\cite{Masip:1996}
%\bibitem{Masip:1996}
%M. Masip and A. Rasin,
  %``CP violation in multi-Higgs supersymmetric models,''
 % Nucl.\ Phys. \ B460 (1996) 449-469
  %doi:	10.1016/0550-3213(95)00600-1
  %[arXiv:9508365 [hep-ph]].


%\cite{Bauer:2015fxa}
%\bibitem{Bauer:2015fxa}
 % M.~Bauer, M.~Carena and K.~Gemmler,
  %``Flavor from the Electroweak Scale,''
  %JHEP {\bf 1511}, 016 (2015)
  %doi:10.1007/JHEP11(2015)016
  %[arXiv:1506.01719 [hep-ph]].
  %%CITATION = doi:10.1007/JHEP11(2015)016;%%
  %8 citations counted in INSPIRE as of 21 Jun 2016


%\cite{Buras:2012dp}
%\bibitem{Buras:2012dp}
 % A.~J.~Buras, F.~De Fazio, J.~Girrbach and M.~V.~Carlucci,
  %``The Anatomy of Quark Flavour Observables in 331 Models in the Flavour Precision Era,''
  %JHEP {\bf 1302}, 023 (2013)
  %doi:10.1007/JHEP02(2013)023
  %[arXiv:1211.1237 [hep-ph]].
  %%CITATION = doi:10.1007/JHEP02(2013)023;%%
  %54 citations counted in INSPIRE as of 21 Jun 2016

\end{thebibliography}
\end{document}